\documentclass[fleqn,usenatbib]{mnras}

\usepackage{newtxtext,newtxmath}

\usepackage[T1]{fontenc}
\usepackage{ae,aecompl}
\usepackage[utf8]{inputenc}

\usepackage{graphicx}	
\usepackage{amsmath}	
\usepackage{amssymb}	
\usepackage[acronym,shortcuts]{glossaries} 
\glsdisablehyper 

\newacronym{DNS}{DNS}{double neutron star}
\newacronym{GW}{GW}{gravitational wave}
\newacronym{BBH}{BBH}{binary black hole}
\newacronym{DCO}{DCO}{double compact object}
\newacronym{DWD}{DWD}{double white dwarf}
\newacronym{ZAMS}{ZAMS}{zero-age main sequence}
\newacronym{MW}{MW}{Milky Way}
\newacronym{SNR}{SNR}{signal-to-noise ratio}
\newacronym{SKA}{SKA}{Square Kilometre Array}
\newacronym[longplural={supernovae}, plural={SNe}]{SN}{SN}{supernova}

\title[Detecting DNSs with LISA]{Detecting Double Neutron Stars with \textit{LISA}}

\author[M. Y. M. Lau et al.]{Mike Y. M. Lau,$^{1,2}$\thanks{E-mail: mike.lau@monash.edu (MYML)}
	Ilya Mandel$^{1,2,3}$,
	Alejandro Vigna-G\'{o}mez$^{4,1,2,3}$,
	\newauthor
	Coenraad J. Neijssel$^{3,1,2}$,
	Simon Stevenson$^{5,2}$,
	and Alberto Sesana$^{6}$
	\\
	$^{1}$ Monash Centre for Astrophysics, School of Physics and Astronomy, Monash University, Clayton, Victoria 3800, Australia\\
	$^{2}$ OzGrav: The ARC Centre of Excellence for Gravitational Wave Discovery, Australia\\
	$^{3}$ Birmingham Institute for Gravitational Wave Astronomy and School of Physics and Astronomy, University of Birmingham, \\Birmingham, B15 2TT, United Kingdom\\
	$^{4}$ DARK, Niels Bohr Institute, University of Copenhagen, Blegdamsvej 17, 2100, Copenhagen, Denmark \\
	$^{5}$ Centre for Astrophysics and Supercomputing, Swinburne University of Technology, Hawthorn, VIC 3122, Australia \\
	$^{6}$ Dipartimento di Fisica ``G. Occhialini'', Universit\'{a} degli Studi Milano Bicocca, Piazza della Scienza 3, I-20126 Milano, Italy
}

\date{Accepted XXX. Received YYY; in original form ZZZ}

\pubyear{2019}

\begin{document}
\label{firstpage}
\pagerange{\pageref{firstpage}--\pageref{lastpage}}
\maketitle

\begin{abstract}
We estimate the properties of the double neutron star (DNS) population that will be observable by the planned space-based interferometer LISA. By following the gravitational radiation driven evolution of DNSs generated from rapid population synthesis of massive binary stars, we estimate that around 35 DNSs will accumulate a signal-to-noise ratio above 8 over a four-year LISA mission. The observed population mainly comprises Galactic DNSs (94 per cent), but detections in the LMC (5 per cent) and SMC (1 per cent) may also be expected. The median orbital frequency of detected DNSs is expected to be 0.8 mHz, and many of them will be eccentric (median eccentricity of $0.11$).  LISA is expected to localise these DNSs to a typical angular resolution of $2^\circ$. We expect the best-constrained DNSs to have eccentricities known to a few parts in a thousand, chirp masses measured to better than 1 per cent fractional uncertainty, and sky localisation at the level of a few arcminutes. The orbital properties will provide insights into DNS progenitors and formation channels. The localisations may allow neutron star natal kick magnitudes to be constrained through the Galactic distribution of DNSs, and make it possible to follow up the sources with radio pulsar searches. LISA is also expected to resolve $\sim 10^4$ Galactic double white dwarfs, many of which may have binary parameters that resemble DNSs; we discuss how the combined measurement of binary eccentricity, chirp mass, and sky location may aid the identification of a DNS. 
\end{abstract}



\section{Introduction}
The LIGO Scientific Collaboration made the first direct detection of \acp{GW} in 2015 from the binary black hole (BBH) merger GW150914 \citep{2016PhRvL.116f1102A}. Since then, eleven \ac{GW} events were recorded in the Gravitational-Wave Transient Catalog (GWTC-1) \citep{GWTC12019}. The start of the Advanced LIGO and Advanced Virgo third observing run (O3) on 1 April, 2019 with improved detector sensitivity has given almost weekly public alerts to credible \ac{GW} candidates on the Gravitational Wave Candidate Event Database (GraceDB). The exploration of \ac{DCO} population statistics will be integral to constraining the relative importance of different formation channels and reducing the large uncertainties that characterise key stages of isolated binary evolution.

\Ac{DNS} coalescences are of particular interest as they may produce electromagnetic counterparts, including gamma ray bursts, their afterglows, and kilonovae. GW170817, detected during the second observing run (O2) of Advanced LIGO and Virgo \citep{TheLIGOScientific:2017qsa}, was associated with EM counterparts GRB 170817A and AT 2017gfo \citep{GW170817:GRB,GW170817:MMA}, marking the first multi-messenger event involving \acp{GW}. 

Neutron stars can receive \ac{SN} natal kicks of several hundred kms${}^{-1}$ and lose significant fractions of mass during \acp{SN}, and so \acp{DNS} forming from isolated binaries may possess significant eccentricities at birth. However, by the time these \acp{DNS} evolve to the 10--1000 Hz sensitivity window of the LIGO-Virgo advanced detectors, gravitational radiation reaction circularises the orbit of isolated binaries to eccentricities $e \lesssim 10^{-5}$ regardless of their formation eccentricity if they are formed at orbital frequencies $\lesssim 10^{-4}$ Hz. On the other hand, ESA's proposed space-based Laser Interferometer Space Antenna (LISA) \citep{Amaro-Seoane+2017} is anticipated to observe inspiral \acp{GW} in the $10^{-4}$--$10^{-2}$ Hz window, and so may detect residual eccentricity in inspiralling \acp{DNS}. Eccentricity measurements by LISA may provide constraints on the physics of isolated binary evolution \citep{Nelemans+2001b,Belczynski+2010,vignagomez2018,Tauris2018,Kyutoku+2019} or dynamical formation \citep{Kremer+2018,Hamers&Thompson2019,Andrews&Mandel2019}.

The detection of GW170817 and the selection of LISA as ESA's third L-class mission in January 2017 has led to recent interest in LISA \acp{DNS} sources. \citet{Seto2019} estimated the frequency distribution of \acp{DNS} in local group galaxies by extrapolating the comoving volumetric \ac{DNS} merger rate inferred from GW170817. \cite{Kyutoku+2019} demonstrated that LISA-informed observations can enhance the efficiency of radio pulsar searches with the \ac{SKA}. \cite{Thrane+2019} showed that constraints may be placed on the neutron star equation of state by measuring the Lense-Thirring precession with multi-messenger observations with LISA and \ac{SKA}.

In this paper, we predict the detection rate, distribution of source parameters (eccentricity, signal-to-noise ratio, distance), and uncertainty in source parameters (eccentricity uncertainty, sky-localisation accuracy, chirp mass uncertainty) of \acp{DNS} in the \ac{MW} and in nearby galaxies. We generate a population of synthetic \acp{DNS} using the \textit{Compact Object Mergers: Population Astrophysics and Statistics} (COMPAS) suite \citep{stevenson2017formation,Barrett+2018,vignagomez2018,Neijssel+2019}, and follow the evolution of these \acp{DNS} through the LISA band driven by gravitational radiation reaction, for which we use the leading quadrupole order expressions \citep{Peters1964}. Starting from an initial population of \ac{ZAMS} binary stars, COMPAS performs single-star evolution using the fitting formulae in \cite{Hurley+2000} and calculates changes in stellar and orbital properties due to wind-driven mass loss, mass transfer, common-envelope events, and \acp{SN}, until the formation of a \ac{DCO}.

This paper is structured as follows. Section \ref{sec:methods} describes how the \ac{DNS} detection rate is calculated; it begins by highlighting important features of COMPAS's \texttt{Fiducial} model of binary evolution (\ref{subsec:popsynth}), then discusses the general procedure for estimating the LISA \ac{DNS} detection rate (\ref{subsec:detrate}), the \ac{DNS} formation rate within the detector's sensitive volume (\ref{subsec:formrate}), and the detector sensitivity (\ref{subsec:SNR}). Section \ref{sec:results} presents our predictions for the distribution of LISA \ac{DNS} binary parameters and their uncertainties. In particular, we discuss how the eccentricity distribution may constrain binary evolution physics. Section \ref{sec:identifyingDNS} discusses strategies for distinguishing LISA \acp{DNS} from resolved Galactic \acp{DWD} and neutron star-white dwarf (NS-WD) binaries. We summarise our results and discuss the validity of our assumptions in Section \ref{sec:conclusions}.

\section{Methods} \label{sec:methods}
\subsection{Population Synthesis} \label{subsec:popsynth}
This work uses a synthetic population of \acp{DNS} evolved by \citet{vignagomez2018} using the rapid population synthesis element of COMPAS. A total of $10^6$ binary stars were evolved, with 0.13 per cent becoming \acp{DNS}, of which 73 per cent merge within the age of the Universe. We highlight distinctive features of the assumed \texttt{Fiducial} model of binary evolution, and refer the reader to \citet{stevenson2017formation} and \citet{vignagomez2018} for details.

The mass of the primary star is drawn from the Kroupa initial mass function \citep{Kroupa2001} in the mass range $[5,100] \ \text{M}_\odot$ (the full mass range was used for normalisation), while the mass ratio $q = m_2/m_1$ is drawn from a uniform distribution in $[0.1,1]$ \citep{Sana+2012}. All binaries are assumed to be circular at \ac{ZAMS} with solar metallicity. The binary separation is drawn from a log-uniform distribution in $[0.1,1000]$ AU, following \cite{Opik1924}.

A common-envelope phase follows dynamically-unstable mass transfer, and is described by the $\alpha\lambda$-formalism \citep{Webbink1984,deKool1990} with $\alpha = 1$ and $\lambda$ determined by the fits of \citet{Xu&Li2010}.

The \texttt{Fiducial} model distinguishes between core-collapse, ultra-stripped, and electron-capture supernovae. The natal kick direction is randomly drawn from the unit sphere while the kick magnitude follows a bimodal distribution. The core-collapse supernova kick magnitude is distributed by a Maxwellian with $\sigma_{\text{high}} = 265$ kms$^{-1}$, following \cite{Hobbs+2005}. Ultra-stripped and electron-capture supernova kicks follow a low-kick Maxwellian with $\sigma_{\text{low}} = 30$ kms$^{-1}$ \citep{Pfahl+2002a,Podsiadlowski+2004,Verbunt+2017}. The `rapid' explosion model in \citet{Fryer2012} is used to calculate the compact remnant mass from the pre-supernova core mass.

A discussion of the \texttt{Fiducial} model's two dominant \ac{DNS} formation channels (accounting for 91 per cent of all \acp{DNS} formed) can be found in \citet{vignagomez2018}. 

To illustrate the sensitivity of our results to uncertainties in binary evolution prescriptions, we compare results obtained with the  \texttt{Fiducial} model assumptions to the following three variants:
\begin{itemize}
	\item Case BB unstable: Case BB mass transfer from a post helium-main-sequence star (see Section \ref{subsubsec:caseBB}) is assumed to always be dynamically unstable, whereas it is always stable in the \texttt{Fiducial} model.
	\item Single \ac{SN} mode: The distribution of natal kick magnitude is a Maxwellian with $\sigma_\text{high} = 265$ kms${}^{-1}$ for all types of \acp{SN}, as opposed to the bimodal distribution in the \texttt{Fiducial} model.
	\item $\alpha = 0.1$: The common-envelope efficiency parameter (see Section \ref{subsubsec:CEalpha}) is set to $\alpha = 0.1$. 
\end{itemize}

We use the \ac{DNS} populations simulated with these variation models\footnote{The case BB unstable, single \ac{SN} mode, and $\alpha = 0.1$ models are the (02), (05), and (10) variations respectively in \cite{vignagomez2018}.} by \citet{vignagomez2018}.

\subsection{Detection Rate} \label{subsec:detrate}
In Monte Carlo population synthesis, each \ac{DNS} synthesised by COMPAS represents a sample population labelled by a set of binary parameters $\bmath{\theta}_i=(e_{0,i}, a_{0,i}, m_{1,i}, m_{2,i})$ for $i=1,2,...,N_\text{DNS}$, where $e_{0,i}$ and $a_{0,i}$ are the eccentricity and semi-major axis at the formation of the $i$th \ac{DNS}, $m_{1,i}$ and $m_{2,i}$ are the component masses, and $N_\text{DNS}$ is the total number of simulated \ac{DNS}. For each \ac{DNS} $\bmath{\theta}_i$, we denote by $f$ its \textit{starting orbital frequency}, the orbital frequency of the \ac{DNS} at the start of its observation by LISA. We also denote by $dN(f) = (dN/df) df$ the number of detections this \ac{DNS} population contributes to the bin $[f,f+df]$ of starting orbital frequencies. The total contribution is therefore
\begin{align}
N = \sum_{i=1}^{N_\text{DNS}} \int_0^\infty \frac{dN_i(f)}{df}df, \label{eq:Ndetsum}
\end{align}
where the subscript $i$ denotes a quantity evaluated for the parameters $\bmath{\theta}_i$. 
The integrand can be written in a more explicit form:
\begin{equation}
N = \sum_{\substack{i=1}}^{N_\text{DNS}} \int_0^\infty \frac{dN_i(t)}{dt} \frac{dt_i(f)}{df} df,
\label{eq:detection_rate}
\end{equation}
which involves (i) $dN_i(t)/dt$, the formation rate of LISA-detectable \acp{DNS} with $\bmath{\theta} = \bmath{\theta}_i$ and (ii) $dt_i(f)$, the time taken for these \acp{DNS} to increase their orbital frequencies from $f$ to $f+df$. The time interval $dt_i$ is calculated by integrating the orbit-averaged, quadrupole-level expression for $[(de/dt)^{-1}](e)$ given in \citet{Peters1964}:
\begin{equation}
	\frac{de}{dt} = -\frac{19}{12} \frac{\beta}{c_0^4}
	\frac{e^{-29/19}(1-e^2)^{3/2}}{[1+(121/304)e^2]^{1181/2299}}
	\label{eq:dedt}
\end{equation}
where $\beta = \frac{64}{5} G^3 m_1 m_2 (m_1 + m_2) / c^5$ and
$c_0 = a_0(1 - e_0^2)  e_0^{-12/19}  (1 + \frac{121}{304} e_0^2)^{-870/ 2299}$ are constants that depend only on the initial binary parameters $\bmath{\theta}_i$. The lower and upper integration limits $e_\text{lower} = e(f)$ and $e_\text{upper} = e(f+df)$, are calculated by inverting the Keplerian expression
\begin{equation}
    f(e) = \frac{1}{2\pi}\sqrt{\frac{G(m_1+m_2)}{a(e)^3}},
\end{equation}
where the orbit-averaged, quadrupole-approximated expression for $a=a(e)$ is also given in \citet{Peters1964}:
\begin{equation}
    a(e) = \frac{c_0 e^{12/19}}{1 - e^2} \bigg(1 + \frac{121}{304} e^2\bigg)^{870/2299}.
	\label{eq:a_e}
\end{equation}

\subsection{DNS Formation Rate} \label{subsec:formrate}
We now discuss how the \ac{DNS} formation rate $dN_i/dt$ is calculated. Since \acp{DNS} are produced by massive stars, their formation rate traces that of massive stars. Significant delay times are possible between star formation and \ac{DNS} merger\footnote{Equation \ref{eq:mergertimeest} gives the merger time of a $m_1=m_2=1.4 \ \text{M}_\odot$ circular \ac{DNS} at the characteristic LISA \ac{GW} frequency $f_\text{GW} = 2$ mHz as $t_\text{merge} =$ 240,000 yr, so LISA typically observes \ac{DNS} shortly before merger relative to the overall \ac{DNS} evolutionary time-scale.}. However, the delay time distribution favours a significant population with short delays, falling off more steeply than $t_\text{delay}^{-1}$ \citep[e.g.,][]{vignagomez2018}. Moreover, our rates are dominated by the \ac{MW}, which does not show evidence of significant star formation rate variations over time.\footnote{The star formation histories of the LMC and SMC show significant variations over time \citep{Harris+Zaritsky2009}, with historical star formation rate up to a factor of 10 lower than the current value, probably leading us to overestimates the \ac{DNS} formation rates in the LMC and SMC.}.

Therefore, we use the \ac{DNS} formation rate as a proxy for the \ac{DNS} merger rate, and use blue light, which traces the massive star formation rate, as a proxy for both \citep[e.g.,][]{Kopparapuetal2008}. To account for long delay times, other approaches that focus on the total mass as a proxy for the \ac{DNS} merger rate are also possible \citep[e.g.,][]{Artale:2019}.

Thus, we take a galaxy's \ac{DNS} formation rate to be proportional to its extinction-corrected blue light luminosity $L_\text{B}$ \citep{Phinney1991,Kalogera+2001,Kopparapuetal2008}. Then, the total formation rate of a \ac{DNS} within some detection volume is proportional to the total blue-light luminosity contained in that volume. As we use a sky-averaged \ac{SNR} in our study, this detection volume is spherical, with radius set by a \ac{SNR} detectability threshold (see Section \ref{subsec:SNR}). With this assumption, the total \ac{DNS} formation rate within distance $d$ is simply the \ac{MW} \ac{DNS} formation rate reweighted by the total blue light luminosity within $d$:
\begin{equation}
	\frac{dN(<d)}{dt} = \frac{L_\text{B}(<d)}{L_\text{B,MW}} \frac{dN_\text{MW}}{dt}.
\end{equation}
The \textit{cumulative blue light luminosity} $L_\text{B}(<d)$ as a function of distance $d$ is derived from the Gravitational Wave Galaxy Catalogue (GWGC) \citep{White2011}, which contains the extinction-corrected absolute blue magnitude of 53,255 galaxies. In particular, we calculate the \ac{MW} blue light luminosity from the GWGC to be $L_\text{B,MW} = 1.07\times 10^{10} \ \text{L}_{B,\odot}$, in units of solar blue light luminosity $\text{L}_{\text{B},\odot}$. For the synthetic \ac{DNS} population in this study, the \ac{MW} \ac{DNS} formation rate\footnote{Table 2 in \citet{vignagomez2018} lists the \textit{merger} rate of Galactic \acp{DNS} to be 24 Myr${}^{-1}$ with a merger fraction of $f_\text{merger}= 0.73$. The total formation rate of Galactic \acp{DNS} is therefore $24 \ \text{Myr}^{-1}/0.73 = 33\text{Myr}^{-1}$.} is $33 \ \text{Myr}^{-1}$ \citep{vignagomez2018}, assuming continuous star formation at $2.0 \ \text{M}_\odot\text{yr}^{-1}$ with solar metallicity $Z_\odot=0.0142$. In Equation \ref{eq:detection_rate}, the formation rate contributed by the $i$th simulated binary is equal to the total formation rate within distance $d_{\text{max},i}$, $dN(<d_{\text{max},i})/dt$, reweighted by the number of simulated \acp{DNS}, $N_\text{DNS}$:
\begin{equation}
	\frac{dN_i}{dt} = \frac{1}{N_\text{DNS}} \frac{dN(<d_\text{max, i})}{dt}.
\label{eq:rate1}
\end{equation}
where $d_{\text{max},i}$ is the \textit{horizon distance} of the $i$th \ac{DNS}, the maximum distance the \ac{DNS} may be located to be detectable by LISA. It is a function of $\bmath{\theta}_i$ and Section \ref{subsec:SNR} explains how it is calculated. This prescription assumes the fraction of massive binary stars that become \acp{DNS} in different galaxies is same as in the \ac{MW}, neglecting variations in, for example, metallicity, binary fraction, and initial mass function. It also neglects variations in star formation rate over cosmic history. Finally, in the \ac{MW}, we focus on \acp{DNS} produced by isolated binary evolution in the Galactic disc and do not consider dynamical formation in globular clusters. The validity of these assumptions is discussed in Section \ref{sec:conclusions}.

We consider two models as limiting cases for the distribution of \acp{DNS} within the Galaxy. The first model assumes negligible kicks or dynamical evolution, so that \acp{DNS} are distributed in the same way as today's massive star birth sites. The second assumes the limit of very large kicks, under which the \acp{DNS} distribution follows the mass distribution of the dark matter halo.

In the first model, we spatially distribute Galactic \acp{DNS} to the plane-projected Galactic disc density profile. We use a disc profile inferred from the disc gravitational potential proposed in \citet{MiyamotoNagai1975},
\begin{equation}
\phi_d(r,z) = -\frac{GM_d}{\sqrt{r^2 + \bigg(a_d+\sqrt{z^2+b_d^2}\bigg)^2}}, 
\end{equation}
where $M_d$ is the total disc mass, $a_d$ and $b_d$ are the radial and vertical scales, and $(r,z)$ are Galactocentric cylindrical coordinates. The density profile $\rho_d(r,z)$ is obtained by solving Poisson's equation $\nabla^2\phi_d = 4\pi G \rho_d$:
\begin{align}
\rho_d(r,z) = \frac{b_d^2M_d}{4\pi} \frac{a_d r^2 + \bigg(a_d + 3\sqrt{z^2 + b_d^2}\bigg)\bigg(a_d + \sqrt{z^2 + b_d^2}\bigg)^2}{\bigg(z^2+b_d^2\bigg)^{3/2}\bigg[r^2 + \bigg(a_d + \sqrt{z^2 + b_d^2}\bigg)^2\bigg]^{5/2}}. \label{eq:discprofile}
\end{align}
Finally, we obtain the plane-projected Galactic disc density profile from the integral $\int \rho_d(r,z) dz$.

In reality, the \ac{DNS} distribution will not trace the birth site distribution because of a combination of natal kicks from asymmetric supernovae, Blaauw kicks produced by symmetric mass loss accompanying supernovae \citep{Blaauw1961}, and subsequent dynamical evolution in the Galaxy's potential. We therefore also consider the opposite extreme: natal kick magnitudes being large enough to eject \acp{DNS} into the dark halo potential. Considering this as a boundary case, we take the extreme limit in which these \acp{DNS} are allowed to relax and virialize, and so trace the dark halo mass distribution. For the \ac{MW} dark halo, we use the density profile in \cite{Wilkinson&Evans1999},
\begin{equation}
	\rho_h(R) = \frac{M_h}{4\pi} \frac{a_h^2}{R^2(R^2+a_h^2)^{3/2}}, \label{eq:DMprofile}
\end{equation}
where $M_h$ is the total halo mass, $a_h$ is a characteristic fall-off radius, and $R$ is the Galactocentric distance. We intentionally do not cut off the halo mass distribution in this model in order to consider it as an extreme limiting case. In Equations \ref{eq:discprofile} and \ref{eq:DMprofile}, we use parameters given in `Model II' of \citet{IrrgangEtAl2013}, obtained by a $\chi^2$-fit to observational constraints: $M_d = 2829 \ \text{M}_\text{gal}$, $a_d = 4.85$ kpc, $b_d=0.184$ kpc, $M_h = 69,725 \ \text{M}_\text{gal}$, $a_h=200$ kpc, and the solar displacement $r_\odot=8.35$ kpc from the Galactic Centre, where $\text{M}_\text{gal} = 2.325 \times 10^7 \ \text{M}_\odot$ is the Galactic mass unit. We expect the true distribution of Galactic \acp{DNS} to be between the Galactic disc and the dark halo scenarios.

\begin{figure}
    \centering
    \includegraphics[width=\columnwidth]{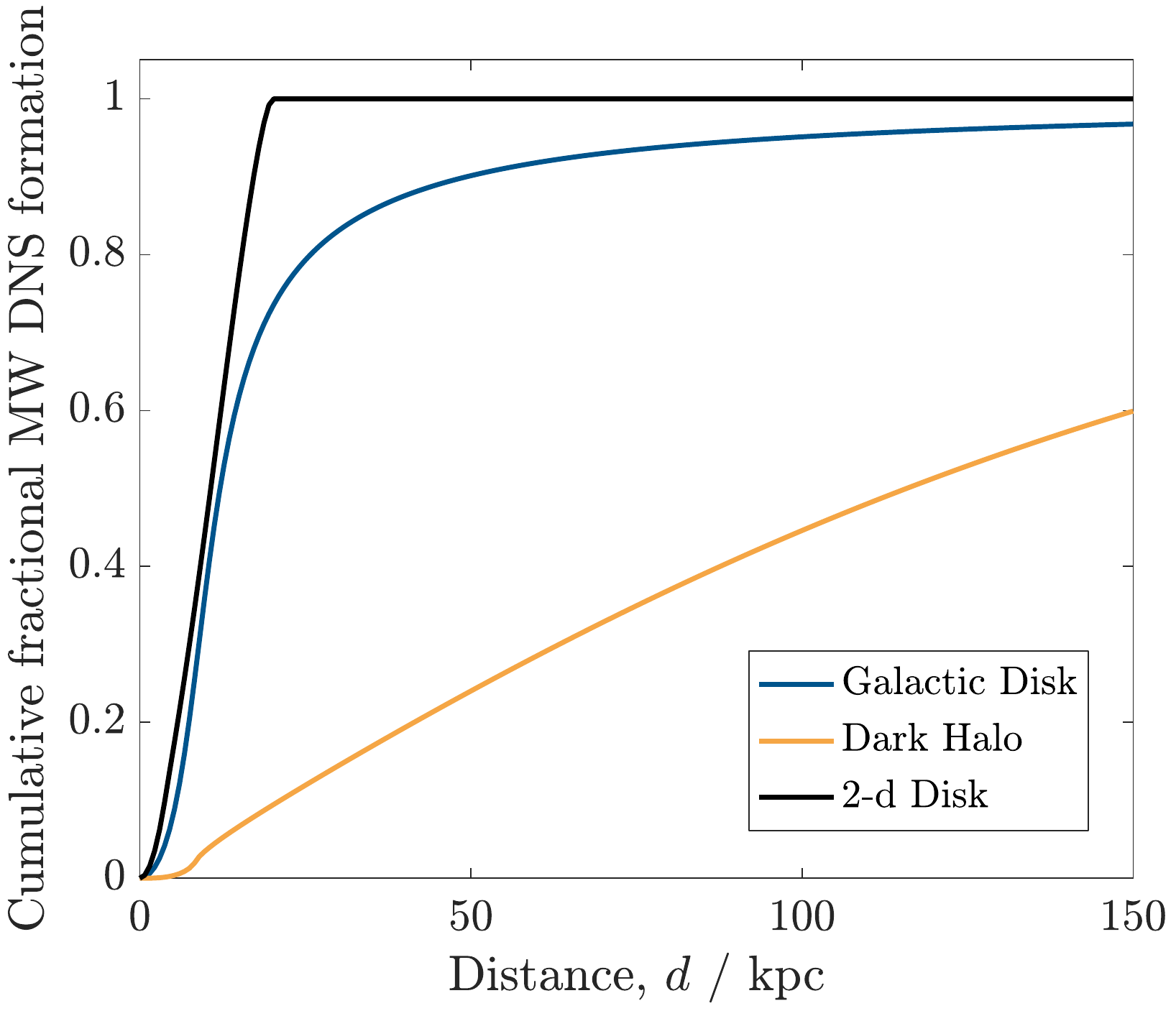}
    \caption{Cumulative fraction of the \ac{MW} \ac{DNS} formation within a given distance from the solar system according to three DNS spatial distribution prescriptions: (i) \ac{DNS} distributed according to the plane-projected Galactic disc density profile (blue); (ii) \ac{DNS} distributed according to the \ac{MW} dark halo density profile (orange); (iii) \ac{DNS} distributed uniformly on a flat disc with radius 12 kpc (black).}
    \label{fig:MW_FR_comparison}
\end{figure}

Figure \ref{fig:MW_FR_comparison} plots the \ac{MW} \ac{DNS} formation rate $dN(<d)/dt$ contained in a spherical detection volume (centred upon the solar system) as a function of the sphere radius $d$ for our two prescriptions. A third, toy prescription has also been included for comparison, where the \ac{MW} is modelled as a uniform flat disc of radius 12 kpc. In the toy model, the detection volume contains the entire disc-like `\ac{MW}' at $d \approx 20$ kpc, beyond which the curve flattens out sharply. The formation rate grows more gently with distance for the Galactic disc potential, where more than 95 per cent of the \ac{DNS} formation rate is contained in $d<100$ kpc. In the dark halo prescription, the diffuse halo stretches out to large distances with scale radius $a_h=200$ kpc, and only 45 per cent of the \ac{DNS} formation rate is contained in $d<100$ kpc. 

\subsection{Signal-to-Noise Ratio} \label{subsec:SNR}
We use an \ac{SNR} expression that is averaged over sky location $(\theta,\phi)$, GW polarisation $\psi$, and source inclination $\iota$. Then, the averaged \ac{SNR} of the $n$th \ac{GW} harmonic depends only on the total energy per unit frequency carried by \acp{GW} emitted in the $n$th harmonic, $dE_n/d(nf)$ \citep[see, e.g.,][]{Flanagan&Hughes1998}\footnote{Note that $\langle \rho_n^2 \rangle$ in Equation \ref{eq:SNR(f)} is larger by a factor of 5 compared to the corresponding expression in LIGO literature, because of the convention in LISA to include the signal response function $\mathcal{R}$ in the noise spectral density as $1/\mathcal{R}_\text{LISA}$, rather than in the strain power spectral density as $\mathcal{R}_\text{LISA}$. Converting from the LIGO to the LISA \ac{SNR} expression requires dividing by a factor of $\mathcal{R}_\text{LIGO} = 1/5$.}:
\begin{equation}
	\langle\rho_n^2\rangle= \frac{2G}{\pi^2c^3d^2} \int_{nf_i}^{nf_f} \frac{|dE_n/d(nf)|}{(nf)^2 \langle S_n(nf) \rangle_{(\theta,\phi)}}d(nf).
	\label{eq:SNR(f)}
\end{equation}
Here, $\langle \rho_n^2 \rangle$ is the squared \ac{SNR} of the $n$th \ac{GW} harmonic averaged over $(\theta,\phi,\iota,\psi)$, $d$ is the source distance, and $\langle S_n(nf) \rangle_{(\theta,\phi)}$ is the sky-averaged LISA one-sided noise power spectral density. The upper and lower integration limits $nf_f$ and $nf_i$ are the $n$th harmonic \ac{GW} frequency of the source at the start and end of LISA observation, respectively. In this study, we assume a four-year LISA mission duration as put forward in the LISA ESA L3 mission proposal \citep{Amaro-Seoane+2017}.

\begin{figure}
	\centering
	\includegraphics[width=\columnwidth]{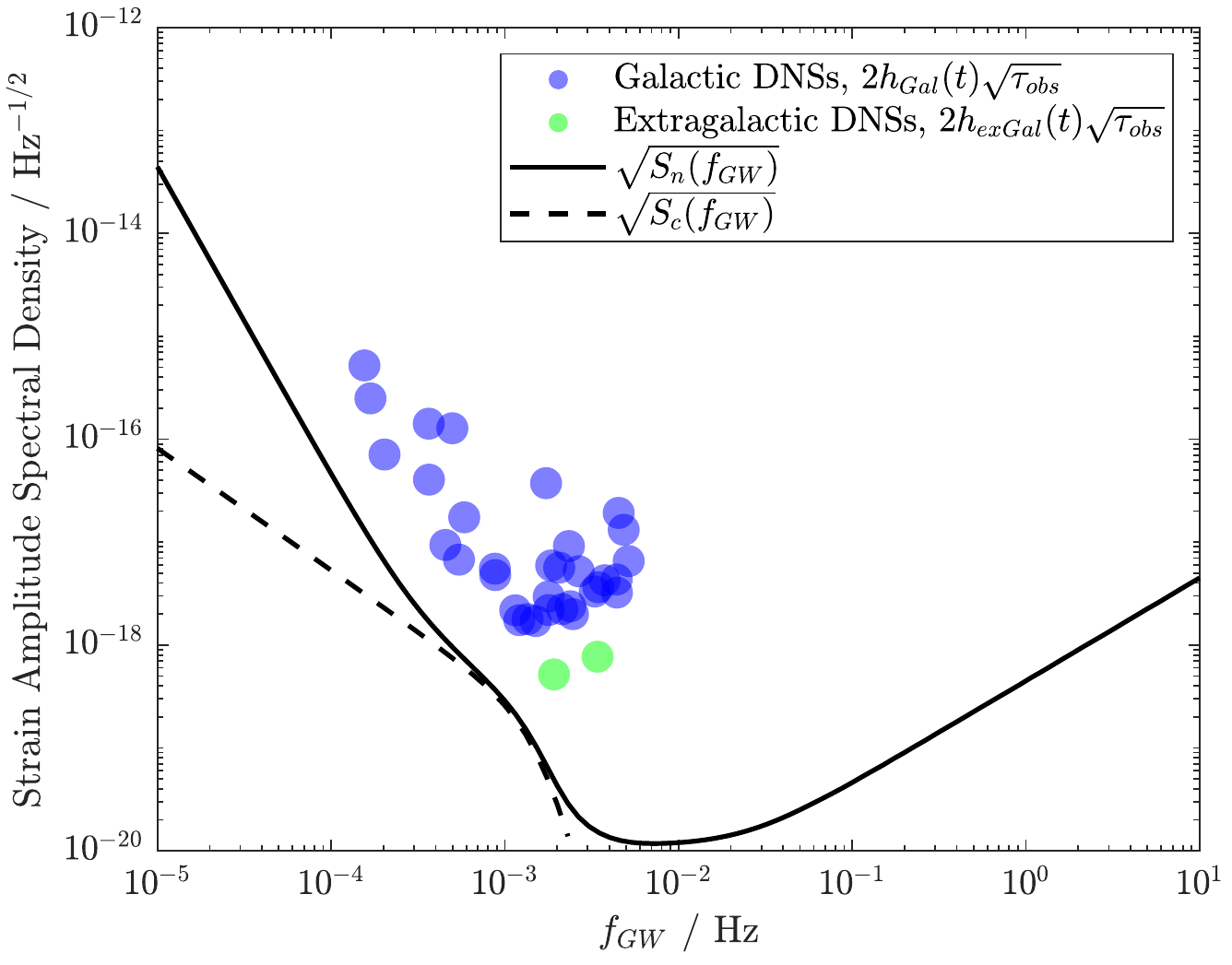}
	\caption{Plot of the total LISA noise amplitude spectral density, $\sqrt{S_n(f_\text{GW})}$ (solid line), the amplitude spectral density of the Galactic background confusion noise, $\sqrt{S_c(f_\text{GW})}$ (dashed line) assuming signal subtraction over a four-year LISA mission, and $2h(t)\sqrt{\tau_\text{obs}}$ for 35 Monte Carlo realisations of LISA \ac{DNS} sources (filled circles) with frequencies drawn from Figure \ref{fig:cumulative_detections} and distances drawn from Figure \ref{fig:distance_dist}. The height of a dot above the solid curve gives the \ac{SNR} of the \ac{DNS}. The green circles correspond to LMC sources.}
	\label{fig:LISAnoiseCurve}
\end{figure}

We approximate the LISA sensitivity curve $S_n(nf)$ with the analytically-fitted expression of \cite{Robson+2019}. This is plotted in Figure \ref{fig:LISAnoiseCurve}, along with a Monte Carlo population of detectable \acp{DNS}. Below \ac{GW} frequencies of 1--3 mHz, the noise spectrum is dominated by confusion noise due to unresolved Galactic binaries, mainly comprising $\sim 10^8$ \acp{DWD} \citep{Nelemans+2001b,Farmer&Phinney2003,Ruiter+2010}. As the LISA mission progresses, the confusion noise reduces since resolved binaries can be removed. We use the set of parameters in \cite{Robson+2019} that assume signal subtraction over a four-year LISA mission. The total \ac{SNR} associated with a (possibly eccentric) source is obtained by summing the \acp{SNR} for each harmonic in quadrature:
\begin{equation}
	\langle \rho^2\rangle = \sum_{n=1}^\infty \langle\rho_n^2\rangle.
	\label{eq:totalSNR}
\end{equation}
In the actual computation, we truncate the sum at the harmonic number
\begin{equation}
	n_\text{cutoff} = \left\| \frac{5\sqrt{1+e}}{(1-e)^{3/2}}  \right\|,
\end{equation}
where $\|k\|$ denotes the nearest integer to $k$. The error in the \ac{GW} luminosity due to this truncation is less than $10^{-3}$ \citep{OLearyetal2009}. We switch to $e$ as the integration variable as $dE_n/dt$ is an explicit function of eccentricity. This is achieved by rewriting in Equation \ref{eq:SNR(f)} $|dE_n/d(nf)| d(nf) = |dE_n/dt||de/dt|^{-1}de$ and substituting the equations for orbit-averaged $|dE_n/dt|$ and $de/dt$ from \cite{PetersMatthews1963} and \cite{Peters1964}:
\begin{align}
	\langle\rho_n^2\rangle= \frac{48}{19} \frac{Gm_1m_2 a_0^2(1-e_0)^2}{c^3d^2M} \int_{e_i}^{e_f} \frac{g(n,e)u(e,e_0)}{n^2 \langle S_n(nf(e)) \rangle_{(\theta,\phi)} }\frac{de}{e}
	\label{eq:SNRe}
\end{align}
where
\begin{equation}
u(e,e_0) = \bigg(\frac{e}{e_0}\bigg)^{\frac{24}{19}} \Bigg(\frac{1+\frac{121}{304}e^2}{1+\frac{121}{304}e_0^2}\Bigg)^{\frac{1740}{2299}} \frac{(1+e_0^2)(1-e^2)^{3/2}}{1-\frac{183}{304}e^2-\frac{121}{304}e^4}
\label{eq:u(e)}
\end{equation}
and $g(n,e)$ determines the relative contribution of each harmonic to the total \ac{GW} luminosity, whose expression is given in \citet{PetersMatthews1963}. The eccentricity at the end of the mission lifetime, $e_f$, is found by integrating Equation \ref{eq:dedt} over the four-year mission lifetime, or set to zero if the binary merges during the mission.

Given the total \ac{SNR} (Equation \ref{eq:totalSNR}) of a \ac{DNS} and a threshold \ac{SNR} $\rho_\text{min}$ for detection, one may calculate the horizon distance $d_\text{max}$ of the source, which is the maximum distance at which this \ac{DNS} is detectable. Using the inverse relationship between the \ac{SNR} $\rho$ and distance $d$ (see Equation \ref{eq:SNRe}), the horizon distance is
\begin{equation}
\frac{d_\text{max}}{\text{kpc}} = \frac{\rho(d = 1 \ \text{kpc})}{\rho_\text{min}}.
\end{equation}
This defines the radius of the spherical detection volume in Equation \ref{eq:rate1} that is required to calculate the formation rate of \acp{DNS}. We assume a four-year LISA mission duration and an \ac{SNR} threshold $\rho_\text{min}=8$ in this study.

\section{Results} \label{sec:results}

\begin{figure}
	\centering
	\includegraphics[width=\columnwidth]{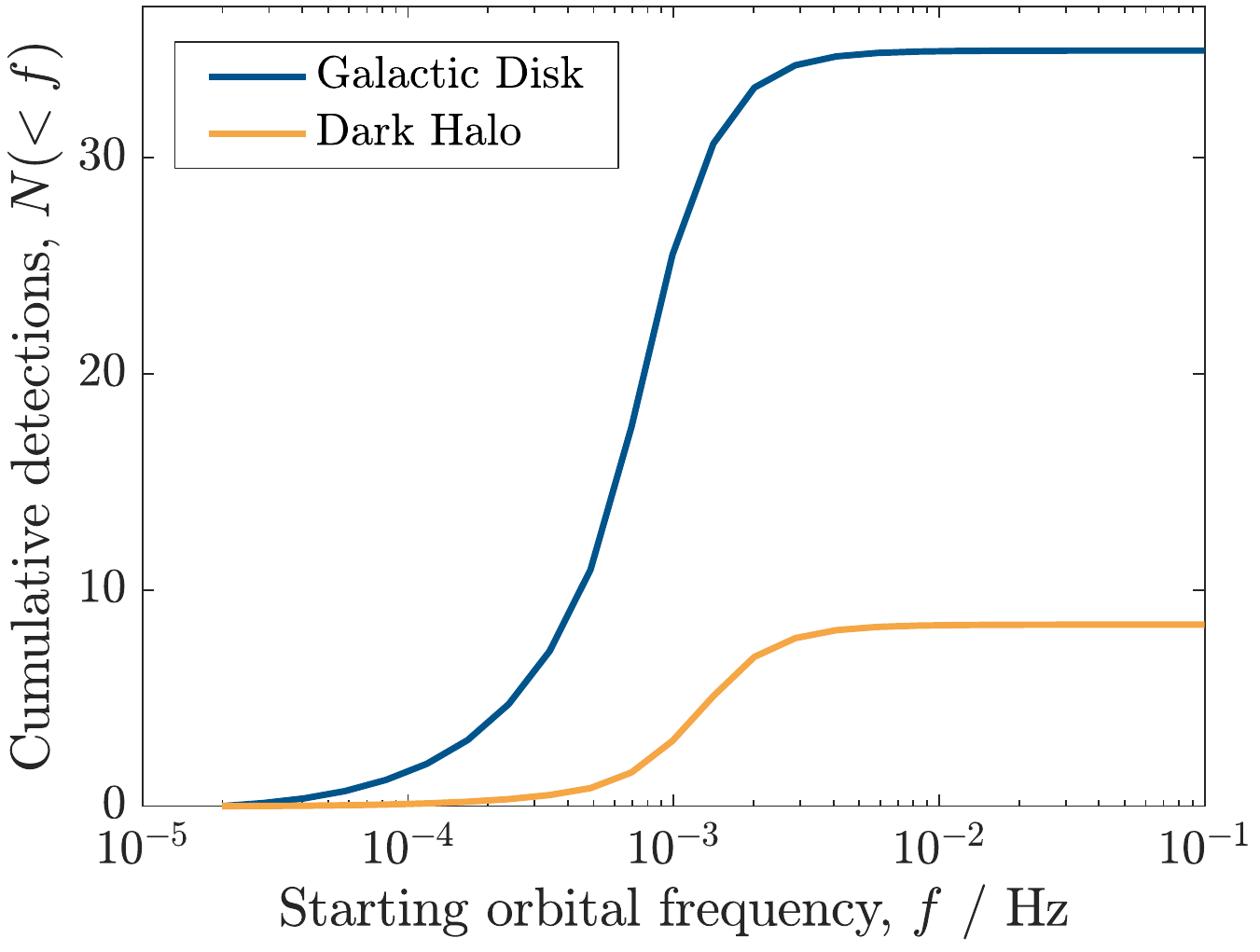}
	\caption{The cumulative number of expected \ac{DNS} detections by LISA over a four-year mission lifetime, as a function of the \ac{DNS} orbital frequency at the start of observation. Blue: Galactic \acp{DNS} distributed according to the plane-projected disc profile. Orange: Galactic \acp{DNS} distributed according to the \ac{MW} dark matter halo density profile.}
	\label{fig:cumulative_detections}
\end{figure}

We anticipate that LISA will be able to detect \acp{GW} from several tens of locals \ac{DNS} binaries. Evaluating Equation \ref{eq:detection_rate} yields 35 detectable \acp{DNS} assuming Galactic sources distributed according to the plane-projected disc density profile (our default for the rest of the paper) and 8.4 \acp{DNS} assuming Galactic sources distributed according to the \ac{MW} dark matter halo density profile. Figure \ref{fig:cumulative_detections} shows the cumulative number of \ac{DNS} detections as a function of the starting orbital frequency $f$. 

\begin{figure}
	\centering
	\includegraphics[width=\columnwidth]{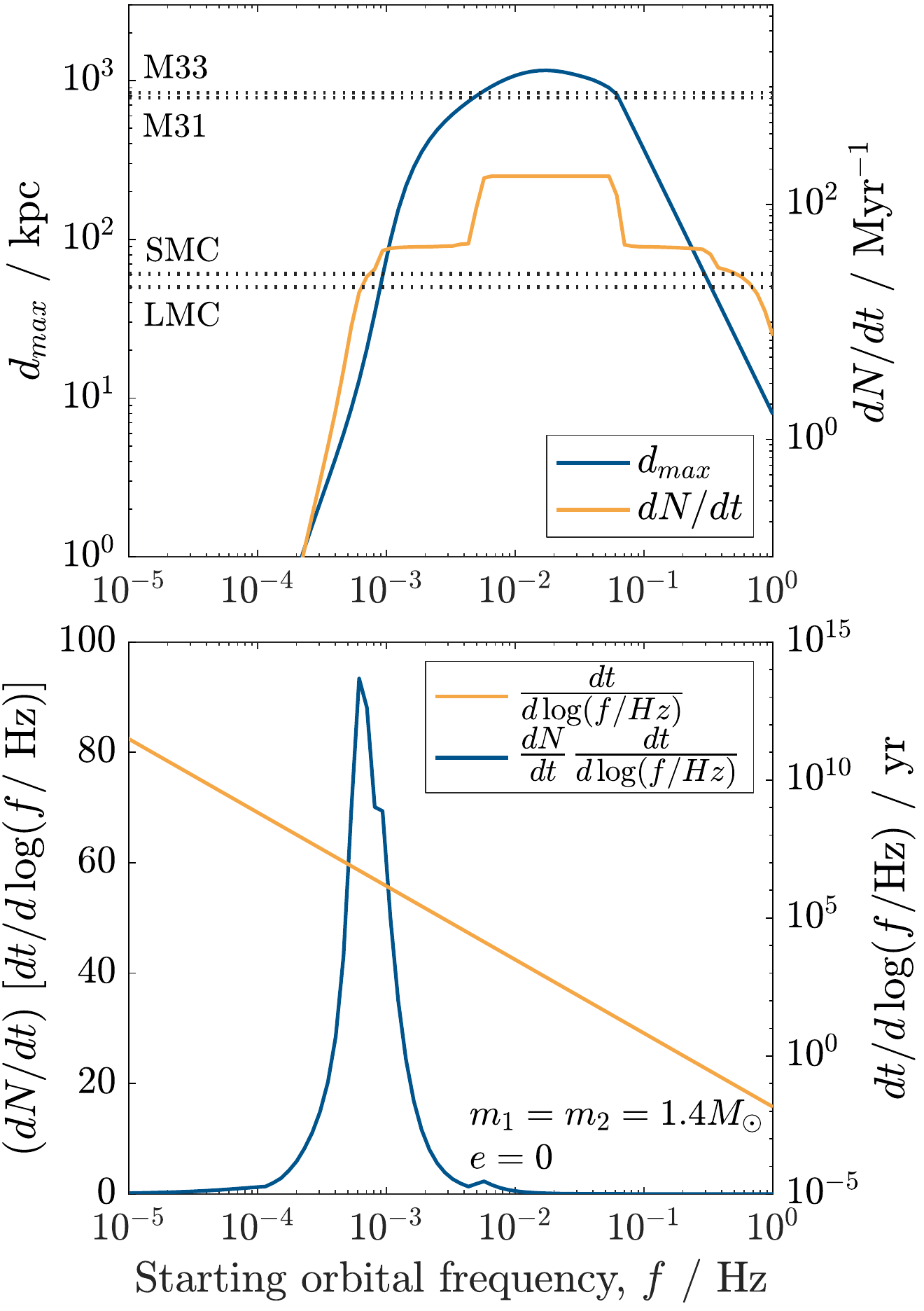}
	\caption{Top: The horizon distance $d_{\text{max}}$ (blue) and the \ac{DNS} formation rate $dN/dt$ within that distance (orange) as functions of starting orbital frequency $f$, with the horizontal dotted lines marking the distances of four nearby galaxies: the LMC, SMC, M31, and M33. Bottom: The \ac{DNS} frequency distribution $(dN/dt)(dt/d\log f)$ (blue), which is the \ac{DNS} formation rate weighted by the time spent by the evolving \ac{DNS} per frequency bin $dt/d\log f$ (orange) as functions of $f$. This figure assumes a circular \ac{DNS} with $m_1=m_2=1.4 \ \text{M}_\odot$.}
	\label{fig:horzVSfreq}
\end{figure}

Although the LISA sensitivity curve is limited by the Galactic confusion noise below \ac{GW} frequencies of 1--3 mHz, LISA \acp{DNS} are detected with 1mHz characteristic orbital frequency (17-minute period), or a gravitational-wave frequency of $2f = 2$ mHz for a circular \ac{DNS}. There are few high-frequency \acp{DNS} in the sample due to the shorter time per unit frequency interval at higher frequencies ($dt/df \propto f^{-11/3}$). This is shown in Figure \ref{fig:horzVSfreq}, which illustrates how our rate estimate is developed using a circular \ac{DNS} with $m_1=m_2=1.4 \ \text{M}_\odot$. The \ac{DNS} can be observed to the greatest distance $d = 1160$ kpc (blue curve in top panel) at 17.5 mHz, yielding the largest \ac{DNS} formation/merger rate within the sensitive volume at that orbital frequency (orange curve in top panel--the apparent steps on this curve correspond to additional galaxies coming into view). However, because of the steep decrease in the time spent by binaries at higher frequencies (orange line in bottom panel), the distribution of expected \ac{DNS} detections per unit logarithmic frequency peaks at 0.6 mHz (blue curve in bottom panel).

\begin{figure}
	\centering
	\includegraphics[width=\columnwidth]{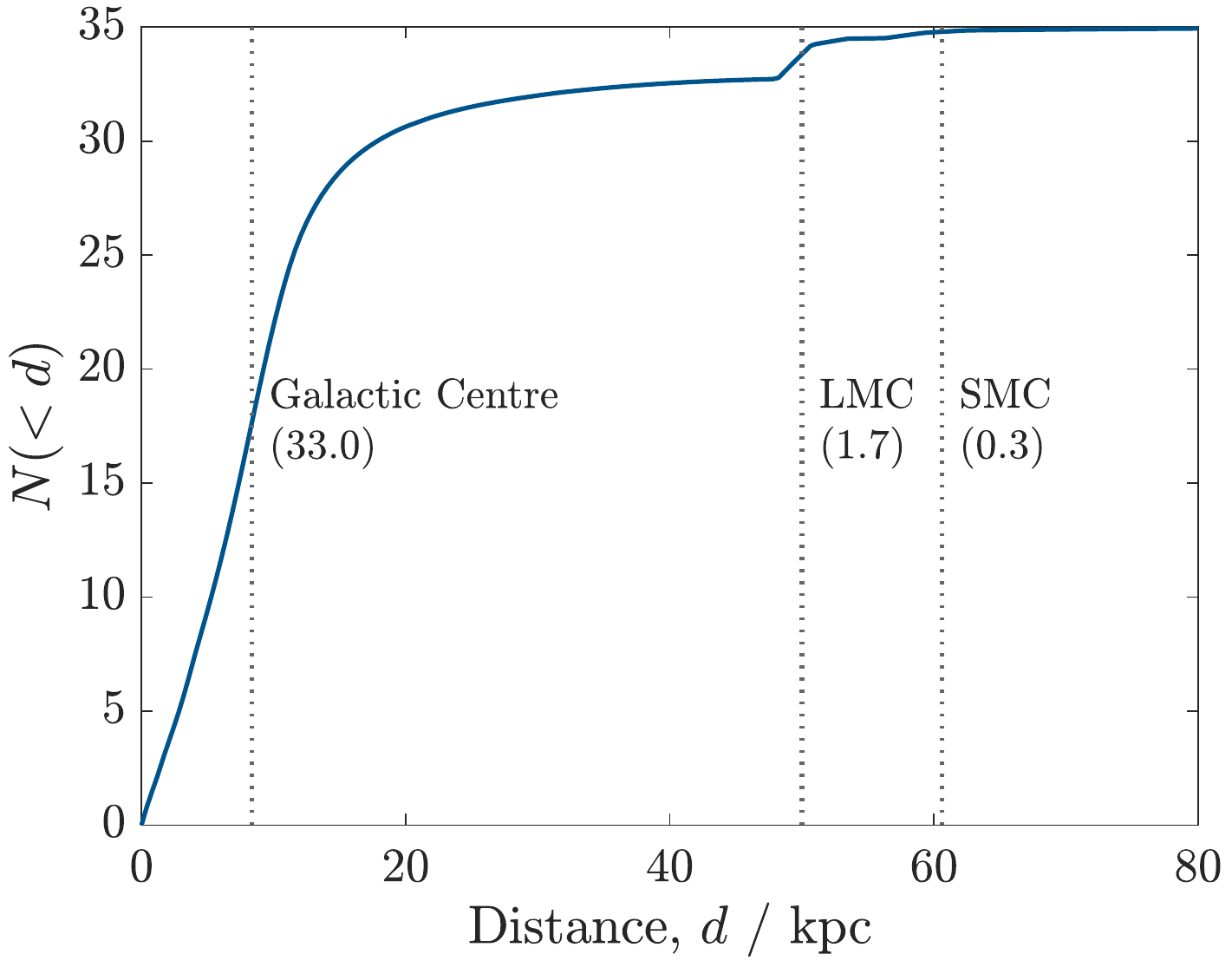}
	\caption{The expected number of LISA \ac{DNS} detections plotted as a function of the distance $d$ from the solar system. The vertical dashed lines mark the positions of the Milky Way nucleus, LMC, and SMC, and the bracketed number gives the number of detections expected in each galaxy.}
	\label{fig:distance_dist}
\end{figure}

Figure \ref{fig:distance_dist} shows the distance distribution of the detectable \acp{DNS}. We predict 33 (94 per cent) of sources to be Galactic, 1.7 (5 per cent) from the LMC and 0.3 (1 per cent) from the SMC. The number of detectable \acp{DNS} in M31 ($d=780$ kpc) and beyond is negligible ($N < 0.01$), as \acp{DNS} spend too little time at orbital frequencies above 10 mHz where they can be observed out to M31 and M33 (see Figure \ref{fig:horzVSfreq}). This is broadly consistent with \cite{Seto2019}, who finds $\sim3$ and $\sim0.5$ detectable \acp{DNS} in the LMC and SMC respectively, using a slightly higher SNR threshold $\rho_\text{min} = 10$ but a larger intrinsic \ac{DNS} merger rate inferred from GW170817. \cite{Seto2019} also expects $\sim1$ detection in M31 and no significant number of detections in M33.

Below, we discuss how accurately \ac{DNS} parameters such as sky location, eccentricity, and chirp mass can be measured from LISA observations. In general, parameter estimation improves with the accumulated \ac{SNR} of a \ac{GW} source, with the typical uncertainty in individual parameters scaling as $1/\rho$ in a regime where the linear signal approximation is valid \citep{CutlerFlanagan:1994,PoissonWill:1995}.

\begin{figure}
	\centering
	\includegraphics[width=\columnwidth]{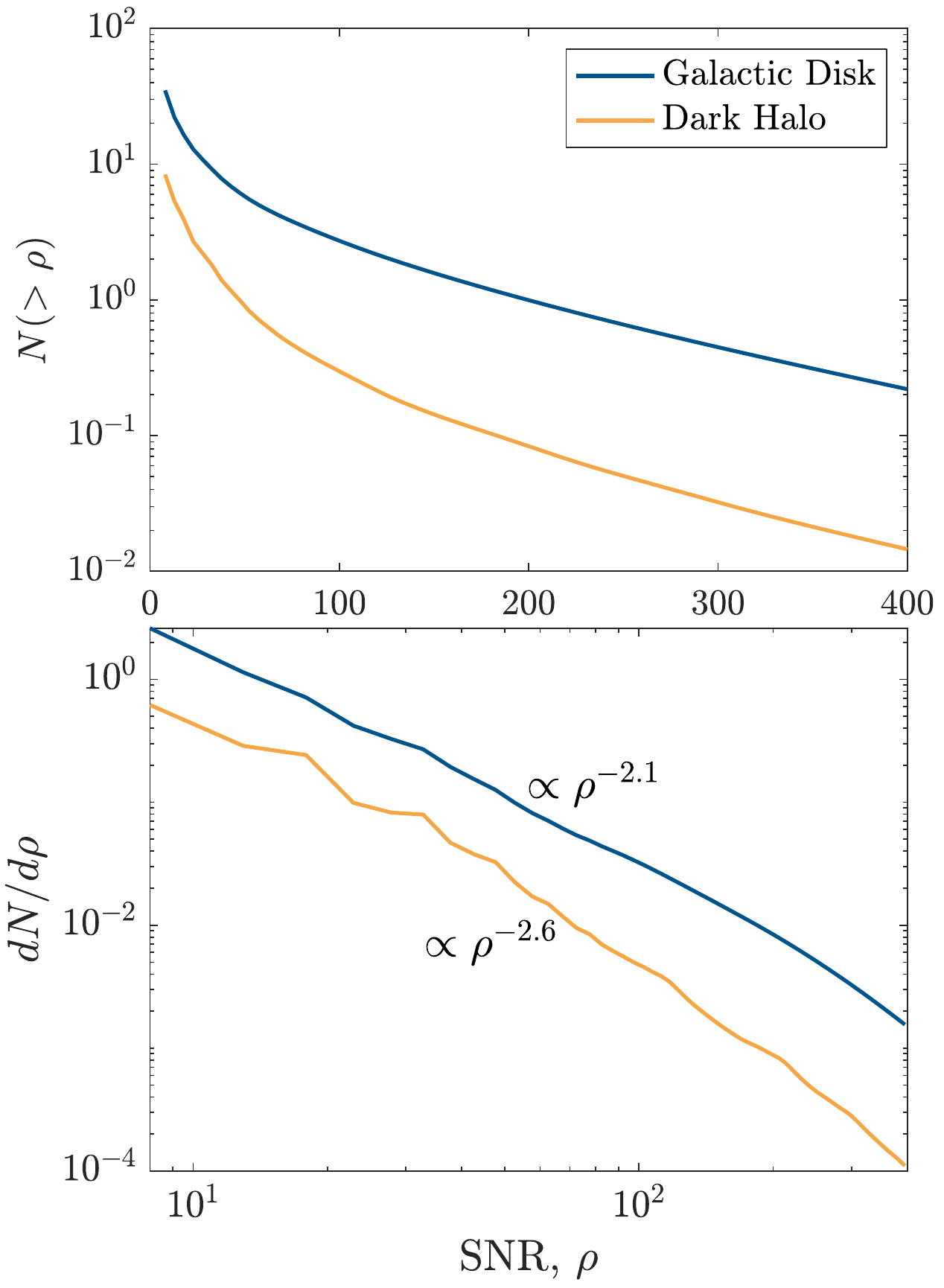}
	\caption{Top: Cumulative \ac{SNR} distribution of detectable \acp{DNS}. Bottom:
		log-log plot of the \ac{SNR} distribution labelled with the power-law index obtained by a least-squares fit.}
	\label{fig:SNR_dist}
\end{figure}

Figure \ref{fig:SNR_dist} shows the cumulative \ac{SNR} distribution of LISA \acp{DNS} for both the disc and the dark matter halo distribution of Galactic \acp{DNS}. For the disc prescription, the median \ac{SNR} is 16.8, $\sim 15$ \acp{DNS} can accumulate $\rho > 20$, 2.5 can accumulate $\rho > 100$, and the highest expected \ac{SNR} (set by $N(>\rho) = 1$) is $\sim 180$. The bottom panel is a log-log plot of $dN/d\rho$ labelled with an approximate slope obtained by a least-squares fit. Because the Galactic disc density is centrally concentrated, $dN/d\rho$ falls off more gently than the expected $\rho^{-3}$ scaling for a uniform disc. Likewise, for the \ac{MW} dark matter halo prescription, $dN/d\rho$ falls off more gently than the expected $\rho^{-4}$ scaling for a uniform sphere. This behaviour also causes the more centrally-concentrated Galactic disc prescription to have a lower characteristic \ac{DNS} frequency in Figure \ref{fig:cumulative_detections}, as closer \acp{DNS} produce stronger signals that can be detected at lower frequencies.

\subsection{Sky Localisation}
Recent works have discussed the importance of sky localisation of LISA \acp{DNS} for multi-messenger follow-ups of Galactic systems, including radio pulsar observations, to constrain the neutron star equation of state and test general relativity \citep{Kyutoku+2019,Thrane+2019}. Accurate sky localisation by LISA can reduce the search time for pulsar surveys such as the \ac{SKA} Phase 2, which may coincide with LISA's expected launch in the 2030s. LISA triggers would thus allow a longer signal accumulation time, which leads to higher detection significance and detections of fainter binary pulsars. \cite{Kyutoku+2019} show that LISA measurements of orbital frequency and other binary parameters can allow a computationally efficient correction of Doppler smearing associated with tight radio pulsars, where the signal integration time is a significant fraction of the orbital period. Moreover, since our calculations show that a total of $\sim 2$ \acp{DNS} may be detected in the LMC and SMC, sky localisation is also needed for host galaxy identification. Three-dimensional localisation is possible if, in addition to sky position, the distance is also well constrained. For Galactic \acp{DNS}, measuring the sky distribution, particularly the displacement from the Galactic plane, will place constraints on \acp{DNS} kick magnitudes. 

\begin{figure}
	\centering
	\includegraphics[width=\columnwidth]{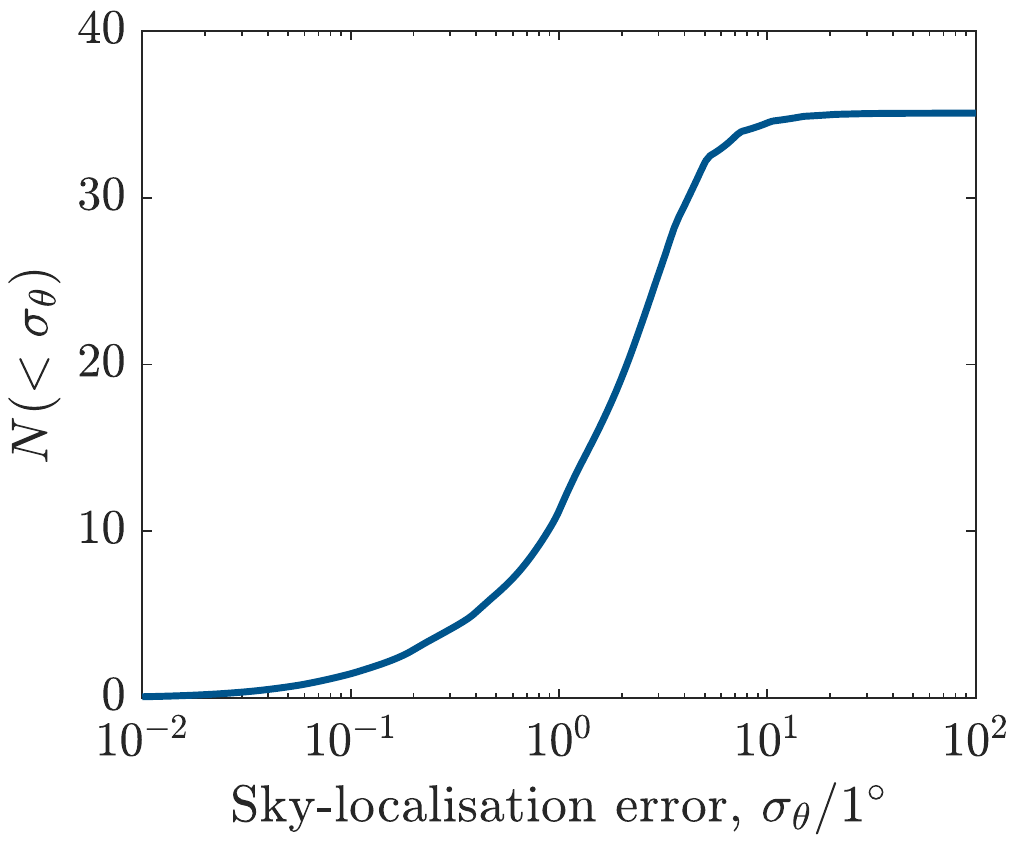}
	\caption{Cumulative distribution of the uncertainty in the sky angle (in one plane) of the simulated detectable \acp{DNS}.}
	\label{fig:skylocalisationdist}
\end{figure}

A \ac{GW} source is triangulated using differences in the signal arrival time in a detector network. For long-lived \ac{GW} sources, this may be accomplished by a single detector observing the source at different points along its orbit around the Sun. The timing error of a \ac{GW} source observed with a detector network is inversely-proportional to the SNR and the detector frequency bandwidth $\sigma_{f_\text{GW}}$ through which the source evolves \citep{Fairhurst2009}. However, LISA \acp{DNS} with orbital frequencies $f \sim 1$ mHz are approximately monochromatic, since their merger time for a circular binary,
\begin{equation}
	\tau_{\text{merge}} = 240,000\ \bigg(\frac{\mathcal{M}_c}{1.2 \ \text{M}_\odot}\bigg)^{-5/3}\bigg(\frac{f_\text{GW}}{0.002 \ \text{Hz}}\bigg)^{-8/3} \ \text{yr}, \label{eq:mergertimeest}
\end{equation}
is much longer than the fiducial four-year LISA mission duration. Then, following \cite{Mandeletal2018}, the timing accuracy instead scales as $1/(\rho f)$, since the \ac{GW} phase is determined down to $1/\rho$ of the wave cycle. LISA will complete multiple heliocentric orbits as it observes a \ac{DNS}, and so gives rise to an effective detector baseline of 2 AU. The uncertainty $\sigma_\theta$ in the source angular coordinate in one plane is approximately
\begin{align}
	\sigma_\theta \approx 2.9 \bigg(\frac{\rho}{10}\bigg)^{-1} \bigg(\frac{f_\text{GW}}{2 \ \text{mHz}}\bigg)^{-1}
	 \bigg(\frac{L}{2\text{AU}}\bigg)^{-1}\text{deg},
	 \label{eq:skylocal}
\end{align}
which is just the timing accuracy divided by the light travel time $L/c$ across the effective detector baseline, and we have used a characteristic \ac{SNR} of 10. This corresponds to localisation within a sky patch of solid angle $\Delta \Omega \sim \pi \sigma_\theta^2 \approx 26.4 \ \text{deg}^2$. We use the approximation of Equation \ref{eq:skylocal} to plot the distribution of $\sigma_\theta$ (Figure \ref{fig:skylocalisationdist}) for our synthetic \ac{DNS} population, finding that most \acp{DNS} can be localised to within $\sigma_\theta \approx 2^\circ$.

An angular resolution of $2^\circ = 0.035$ rad allows the vertical displacement of a Galactic \ac{DNS} above the Galactic plane at $d=10$ kpc to be resolved to $(0.035 \ \text{rad})(10 \ \text{kpc})=0.35$ kpc, roughly the thickness of the old thin stellar disc itself. The size of a pencil beam for a 15 m diameter \ac{SKA} dish observing at 1.4 GHz is approximately $0.67 \ \text{deg}^2$ \citep{Smits+2009,Kyutoku+2019}. It then follows from Figure \ref{fig:skylocalisationdist} that $\approx 6$ \acp{DNS}, if containing a radio pulsar, can be covered by a single pointing with the \ac{SKA}.

\subsection{Eccentricity} \label{subsec:ecc}
Significant orbital eccentricities may be imparted to \acp{DNS} by supernova kicks \citep[e.g.,][]{Tauris+2017} or Blaauw kicks \citep{Blaauw1961}. Short-period \acp{DNS} may also be formed through dynamical hardening interactions in globular clusters until the binary is ejected into the field, presents too small of a cross-section for further interactions, or merges through \ac{GW} emission \citep{Kulkarni+1990,Phinney&Sigurdsson1991}, or in hierarchical triple-star systems \citep{Hamers&Thompson2019}. The typical separation of the ejected \acp{DNS} depends on the globular cluster properties, but may fall in the range of a few solar radii, or orbital frequencies of a few times $10^{-5}$ Hz \citep{Andrews&Mandel2019}. These ejected systems sample a thermal eccentricity distribution $p(e) = 2e$ \citep{Heggie1975}, thereby producing high-frequency, eccentric \ac{GW} sources. However, \acp{DCO} typically circularise before reaching the 10-1000 Hz \ac{GW} sensitivity window of the Advanced LIGO and Virgo detector networks due to gravitational radiation reaction \citep{Peters1964}, though some dynamical channels may yield observable eccentricities in the ground-based detector frequency band \citep[e.g.,][]{Samsing:2014}. On the other hand, even field \acp{DNS} possess measurable residual eccentricities in LISA's millihertz \ac{GW} window, giving important insights into \ac{DNS} formation channels and their progenitor properties. 

The blue solid line of Figure \ref{fig:eccdist} shows the expected eccentricity distribution of the \ac{DNS} population observed by LISA, assuming the isolated binary evolution channel as predicted by the COMPAS \texttt{Fiducial} model of \citet{vignagomez2018}. We find that this model predicts a significant number of eccentric \acp{DNS} in the LISA band with median eccentricity of 0.11 at detection and several highly eccentric systems, e.g. $N(e > 0.6) = 3.6$. To illustrate how binary physics may be imprinted onto the LISA \ac{DNS} distribution, we also include the eccentricity distributions of \acp{DNS} simulated with variations in binary evolution prescription.

\subsubsection{Case BB Mass Transfer Stability} \label{subsubsec:caseBB}
Case BB mass transfer refers to Roche lobe overflow from a post helium-main-sequence star (a helium Hertzsprung-gap star) \citep[e.g.,][]{Delgado&Thomas1981,Ivanova:2003}. In the \ac{DNS} formation channels identified by \cite{vignagomez2018}, this is initiated by a secondary that has previously been stripped of its hydrogen envelope during a common-envelope event. Case BB mass transfer leads to further stripping of the helium envelope down to a metal core, resulting in an `ultra-stripped' star \citep{Tauris+2013,Tauris+2015}. This stripping may leave a thin carbon and helium layer, which allows the ensuing ultra-stripped \ac{SN} to receive a low but non-zero supernova natal kick, along with the Blaauw kick from symmetric mass loss. This allows the \ac{DNS} to become eccentric despite previously going through a common-envelope.
	
The COMPAS \texttt{Fiducial} model assumes that case BB mass transfer is always stable, which is justified \textit{a posteriori} by the better match to the observed Galactic \ac{DNS} period--eccentricity distribution.  Moreover, all simulated systems undergoing case BB mass transfer meet the mass ratio-period stability criterion of \cite{Tauris+2015} and more than 90 per cent meet the mass ratio stability criterion of \cite{Claeys+2014}.
	
The orange dashed curve of Figure \ref{fig:eccdist} shows the eccentricity distribution of \acp{DNS} detectable by LISA under the assumption that case BB mass transfer is always dynamically unstable instead. With this model variation, case BB mass transfer always leads to a common-envelope phase, which significantly tightens the orbit and produces \acp{DNS} with $\sim 1$ mHz orbital frequencies. This is right in the detectability region of LISA, and so these \acp{DNS} undergo little to no circularisation by gravitational radiation by the time they are detected. This is reflected by the higher median eccentricity of 0.36. However, unstable case BB mass transfer leads to fewer overall detections (see Appendix \ref{app:freq_dist_models}).  Although the total \ac{DNS} merger rate in the unstable case BB variation is similar to that in the \texttt{Fiducial} model with stable case BB mass transfer, unstable case BB produces tighter, higher-frequency binaries that evolve rapidly through the LISA sensitive frequency window, leading to fewer observable systems at a given time.  Both stable and unstable case BB mass transfer could occur in reality, so the \texttt{Fiducial} (blue solid curve) and case BB unstable (orange dashed curve) models represent boundary cases.

\subsubsection{Natal Kick Magnitude Distribution}
The distribution of neutron star natal kicks is another uncertainty in binary population synthesis. \cite{Hobbs+2005} proposed a Maxwellian distribution with scale parameter $\sigma = 265$ kms${}^{-1}$ based on the observed 2-d pulsar velocity distribution, while \cite{Verbunt+2017} suggest that a bimodal Maxwellian produces a better agreement because it better fits the low-speed pulsar subpopulation. Population synthesis studies also suggest that the bimodal distribution is needed to match the observed wide Galactic \acp{DNS}, which are overwhelmingly disrupted by a natal kick drawn from a single, high-velocity mode.

Figure \ref{fig:eccdist} shows that the single high \ac{SN} natal kick variation (dotted purple curve) produces a moderately more eccentric population than the \texttt{Fiducial} bimodal distribution, with median $e=0.23$.  However, the most significant difference relative to the \texttt{Fiducial} model is an overall decrease in the number of detectable \ac{DNS} systems by almost a factor of 3, as more binaries are disrupted by the greater \ac{SN} natal kicks (see Figure \ref{fig:cumulativeDetectionsModelComparison}).

\subsubsection{Common-Envelope Efficiency}
\label{subsubsec:CEalpha}
\begin{figure}
	\centering
	\includegraphics[width=\columnwidth]{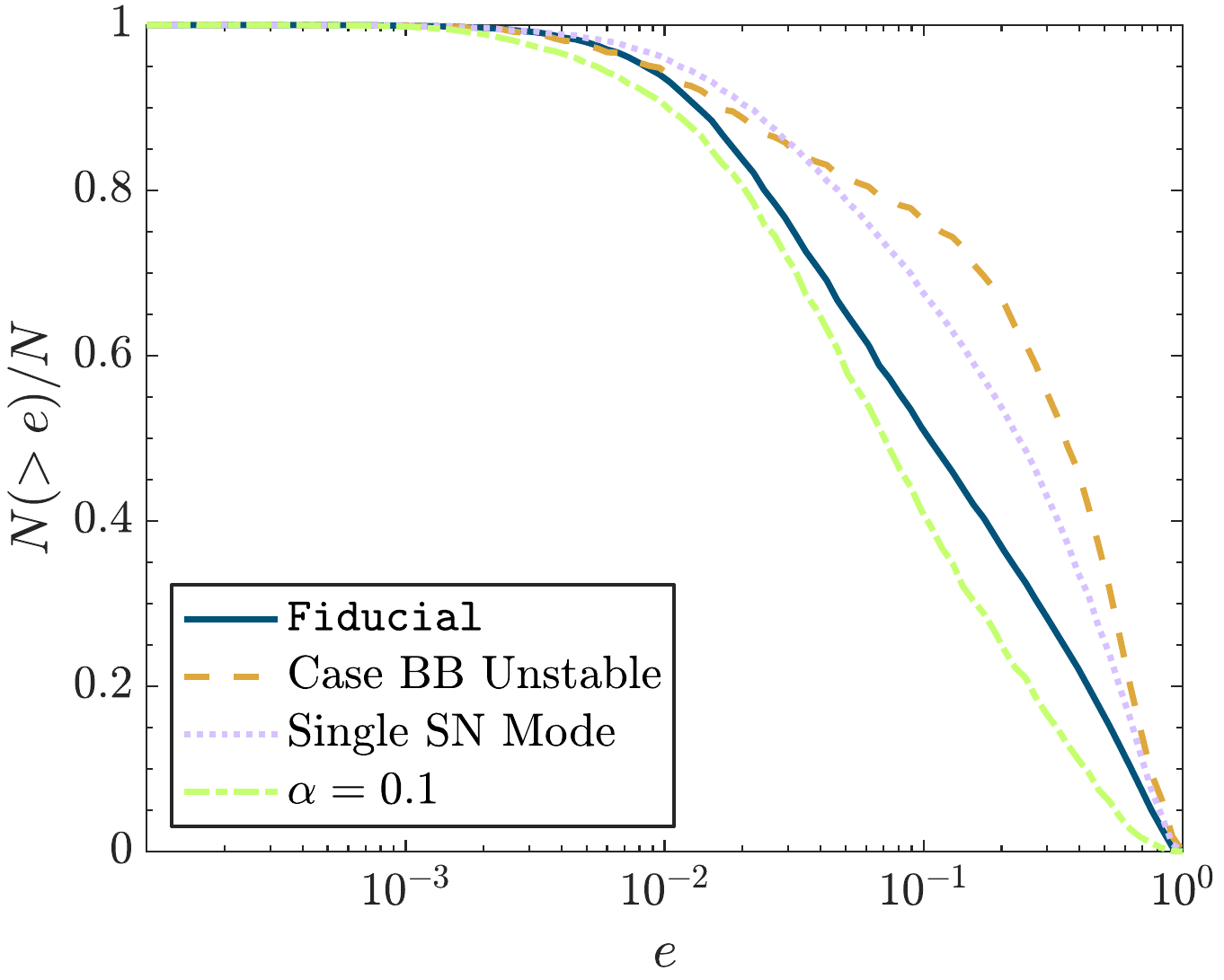}
	\caption{Normalised cumulative eccentricity distribution for LISA-detectable \acp{DNS} for the COMPAS \texttt{Fiducial} model (blue solid curve) of binary evolution, and for variations with: always dynamically unstable case BB mass transfer (orange dashed), a single \ac{SN} kick magnitude (purple dotted), and a common-envelope efficiency of $\alpha = 0.1$ (green dash-dot).}
	\label{fig:eccdist}
\end{figure}

The common-envelope efficiency parameter $\alpha$ \citep{Webbink1984,deKool1990} is the ratio of the binding energy of the common envelope to the difference in orbital energy before and after the common-envelope phase. The \texttt{Fiducial} model's default value of $\alpha = 1$ assumes perfectly efficient transfer of orbital energy into unbinding the envelope, while $0 < \alpha < 1$ assumes that this energy transfer is not fully efficient. We consider a variation with $\alpha = 0.1$.  The green dash-dot curve of Figure \ref{fig:eccdist} shows the corresponding eccentricity distribution, which has a moderately less eccentric population than the \texttt{Fiducial} model, with a median eccentricity of 0.071.

The examples above highlight the value of LISA eccentricity measurements to constraining the physics of binary evolution. Figure \ref{fig:cumulativeDetectionsModelComparison} shows that the same model variations do not significantly affect the frequency distribution of \acp{DNS} at the moment of detection by LISA, which is mainly driven by the LISA sensitivity; it also highlights the differences in the overall rates between variations.

\subsubsection{Eccentricity Measurement}
We consider a conservative threshold for the detection of multiple harmonics by testing whether individual harmonics pass the \ac{SNR} detection threshold \citep[e.g.][]{Willems+2007}; in practice, this condition may be relaxed with the aid of a matched-filtering search for eccentric signals. The uncertainty in measured eccentricity depends strongly on whether two or more harmonics are individually detected, or only a single harmonic is detected, and so we consider these cases separately.

\begin{figure}
	\centering
	\includegraphics[width=\columnwidth]{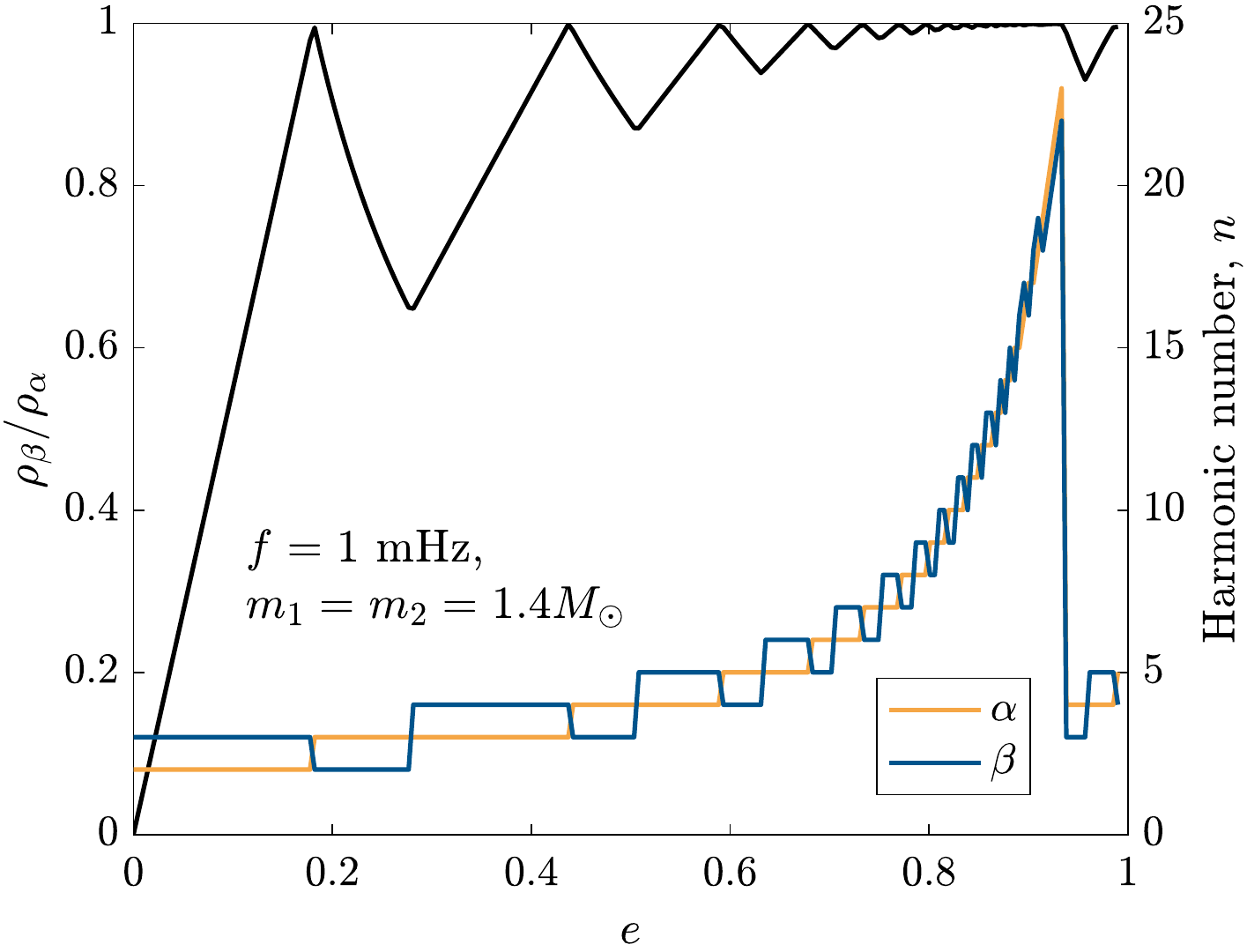}
	\caption{The \ac{SNR} ratio of the second-loudest to the most loudest \ac{GW} harmonic, $\rho_\beta/\rho_\alpha$ (black), as a function of eccentricity for a DNS with $m_1 = m_2 = 1.4 \ \text{M}_\odot$ with starting orbital frequency $f = 1$ mHz. Also plotted are the harmonic numbers $n$ corresponding to the loudest \ac{GW} harmonic, $\alpha$ (orange), and second-loudest \ac{GW} harmonic, $\beta$ (blue).}
	\label{fig:SNRratio}
\end{figure}

If two or more harmonics are detected, the source eccentricity can be determined from the ratio of \ac{GW} amplitudes of these harmonics. We denote the \ac{SNR} and harmonic number of the loudest (largest \ac{SNR}) harmonic by $\rho_\alpha$ and $\alpha$ respectively, and denote the respective quantities for the second-loudest harmonic by $\rho_\beta$ and $\beta$. The harmonic numbers $\alpha$ and $\beta$ and the orbital frequency $f$ can be determined from the observed harmonic frequencies $\alpha f$ and $\beta f$ with the additional knowledge that the two loudest harmonics are neighbouring, $\beta = \alpha \pm 1$. Then the \ac{SNR} ratio $\rho_\beta/\rho_\alpha \in (0,1)$ can be mapped uniquely to the source eccentricity $e$. We plot $\rho_\beta/\rho_\alpha$ as a function of $e$ in Figure \ref{fig:SNRratio} for a typical \ac{DNS} ($f=1$ mHz and $m_1=m_2=1.4 \ \text{M}_\odot$) observed by LISA. The upward trend in $\rho_\beta/\rho_\alpha$ with increasing eccentricity reflects the dispersal of \ac{GW} luminosity across a larger range of frequency harmonics for a more eccentric source. However, there are also spiked structures in the plot originating from $\alpha$ and $\beta$ interchanging values as a \ac{DNS}'s eccentricity decreases. 

In Figure \ref{fig:SNRratio}, $\alpha$ and $\beta$ drop abruptly at $e=0.93$ to $\alpha = 4$ and $\beta = 3$. This occurs because although the peak \ac{GW} luminosity shifts to higher harmonics as eccentricity increases, it is also emitted at increasingly larger frequencies away from the trough of the LISA noise curve and so is suppressed. For the 1 mHz orbital frequency chosen for this example, this suppression becomes sufficiently large at $e = 0.93$ that the $n=4$ harmonic becomes loudest because its frequency falls in the region of minimum noise, $\sim4$ mHz. This shows that very eccentric \acp{DNS} may be detected as systems with dominant harmonics that have similar \acp{SNR} but small harmonic numbers. 

Meanwhile, for a \ac{DNS} with only one detectable \ac{GW} harmonic, an upper constraint may be placed on the eccentricity based on the fact that the harmonic with the second-largest \ac{SNR} is below the detection threshold: $\rho_\beta < \rho_\text{min} \implies \rho_\beta/\rho_\alpha < \rho_\text{min}/\rho_\alpha$. This maximum \ac{SNR} ratio can then be mapped to a maximum eccentricity. For example, in Figure \ref{fig:SNRratio}, constraining the eccentricity to $e < 0.1$ requires $\rho_\beta/\rho_\alpha < 0.5$, i.e., the \ac{SNR} in the $n=2$ harmonic would need to be at least a factor of two above the detection threshold. Therefore, eccentricity is relatively poorly constrained for \acp{DNS} with only one detectable harmonic.

The uncertainty in measured eccentricity $e$ is further compounded by fluctuations in the \ac{SNR} due to noise. While Figure \ref{fig:SNRratio} shows the ratio of expected \acp{SNR}, actual \acp{SNR} fluctuate at the level of $\pm 1$ for different noise realisations. Consequently, in the limit of large \ac{SNR}, the uncertainty on the \ac{SNR} ratio $\rho_\beta / \rho_\alpha$ is at the level $1/\rho_\beta + 1/\rho_\alpha$.

\begin{figure}
	\centering
	\includegraphics[width=\columnwidth]{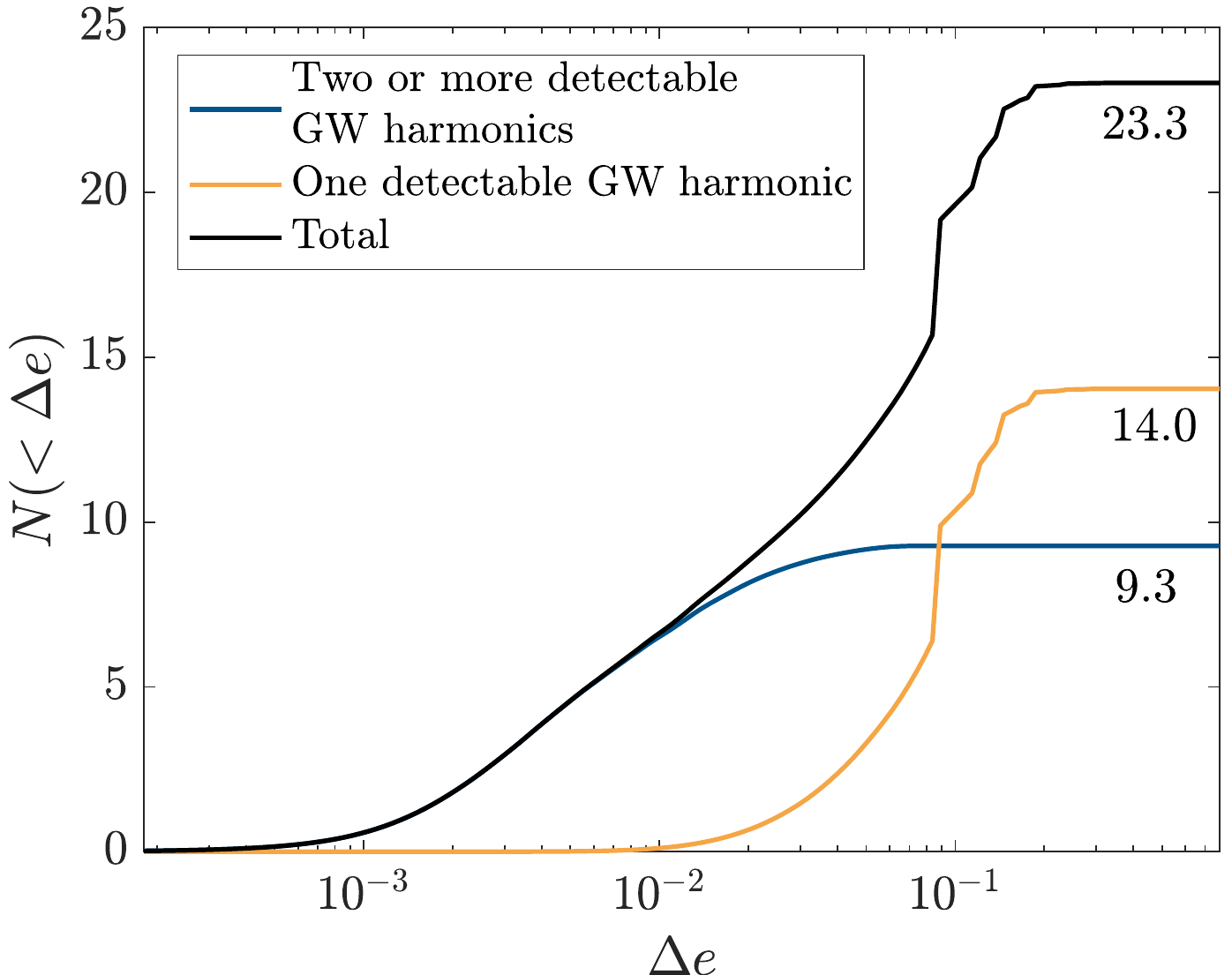}
	\caption{Cumulative distribution of eccentricity uncertainty for LISA DNSs with one detectable GW harmonic (orange) and with two or more detectable GW harmonics (blue). The total distribution (black) contains fewer than 35 systems because for some systems the total \ac{SNR} exceeds the detection threshold but no individual harmonics do so.}
	\label{fig:ecc_err}
\end{figure}

Figure \ref{fig:ecc_err} shows the distribution of eccentricity uncertainties based on $\rho_\beta/\rho_\alpha$ vs. $e$ such as Figure \ref{fig:SNRratio} for each starting \ac{DNS} frequency. We find that there are 9 (26 per cent) \acp{DNS} with two or more detectable harmonics, for which eccentricity is determined to within a few times $10^{-3}$ to a few times $10^{-2}$, and 14 (40 per cent) DNSs with only one detectable harmonic, for which eccentricity is determined to within 0.1--0.2. The remaining 11.7 \acp{DNS} (33 per cent) pass the total \ac{SNR} detection threshold (Equation \ref{eq:totalSNR}) but without any individually detectable harmonics.

Measuring the eccentricity distribution would provide an important probe of binary evolution physics, e.g., distinguishing between the two models shown in Figure \ref{fig:eccdist}. 

\subsection{Mass Measurement} \label{subsec:mass}
For circular binaries, the chirp mass $\mathcal{M}_c = m_1^{3/5} m_2^{3/5} (m_1+m_2)^{-1/5}$ can be directly inferred from the frequency and its rate of evolution in time. For an eccentric binary, the frequency evolution depends on both the chirp mass and the eccentricity:
\begin{equation}
n\dot{f}(\mathcal{M}_c,f,e) = 
\frac{96}{5} \bigg(\frac{2\pi}{n}\bigg)^{8/3} (nf)^{11/3} 
\bigg( \frac{G\mathcal{M}_c}{c^3} \bigg)^{5/3}F(e),
\label{eq:fdot}
\end{equation}
where
\begin{align}
F(e) = \frac{1+\frac{73}{24}e^2 + \frac{37}{96}e^4}{(1-e^2)^{7/2}} 
\label{eq:enhancementfactor}
\end{align}
is the enhancement factor, and setting $n=1$ gives the expression for the \textit{orbital frequency chirp}, $\dot{f}$. Therefore, the imprints of the eccentricity and chirp mass are correlated and they must be measured simultaneously, although the limit $F(e) \geq 1$ on the enhancement factor implies that an upper limit on the chirp mass can be safely obtained by setting $F(e)=1$.

Once $f$, $\dot{f}$, and $e$ are measured from the \ac{GW} signal, the chirp mass $\mathcal{M}_c$ may be determined from Equation \ref{eq:fdot}. It also follows from Equation \ref{eq:fdot} that the fractional uncertainty in chirp mass is
\begin{equation}
    \frac{\Delta \mathcal{M}_c}{\mathcal{M}_c} = 
     \frac{11}{5}\frac{\Delta f}{f} + \frac{3}{5} \frac{\Delta \dot{f}}{\dot{f}} + \frac{3}{5} \frac{\Delta F(e)}{F(e)}. \label{eq:chirpmasserror}
\end{equation}
For a \ac{DNS} that is observed over time $\tau_\text{obs}$ by LISA and has \ac{SNR} $\rho$, the uncertainties in $f$ and $\dot{f}$ are $\Delta f \approx 2.2/(\rho\tau_\text{obs})$ and $\Delta \dot{f} \approx 4.3/(\rho\tau_\text{obs}^2)$ \citep{TakahashiSeto:2002}. From this and using Equation \ref{eq:fdot} for $\dot{f}$, we have the scalings
\begin{align}
\frac{\Delta f}{f} &= 8.7 \times 10^{-7} \ \bigg(\frac{f_\text{GW}}{2 \ \text{mHz}} \bigg)^{-1} \bigg(\frac{\rho}{10}\bigg)^{-1} \bigg(\frac{\tau_\text{obs}}{4 \ \text{yr}}\bigg)^{-1}\ , \\
\frac{\Delta \dot{f}}{\dot{f}} &= 0.26 \ \bigg(\frac{f_\text{GW}}{2 \ \text{mHz}} \bigg)^{-11/3} \bigg(\frac{\rho}{10}\bigg)^{-1} \bigg(\frac{\tau_\text{obs}}{4 \ \text{yr}}\bigg)^{-2} \bigg(\frac{\mathcal{M}_c}{1.2 \ \text{M}_\odot}\bigg)^{-5/3} \label{eq:Deltafdot}
\end{align}
for a circular \acp{DNS}. This suggests that the contribution of the frequency measurement uncertainty to the chirp mass measurement uncertainty can be neglected. The contribution to chirp mass error due to eccentricity, $\Delta F(e)$, can be calculated directly for known $\Delta e$ using Equation \ref{eq:enhancementfactor} for $F(e)$, while the uncertainty in $e$ may be calculated as described in Section \ref{subsec:ecc}. 

\begin{figure}
	\centering
	\includegraphics[width=\columnwidth]{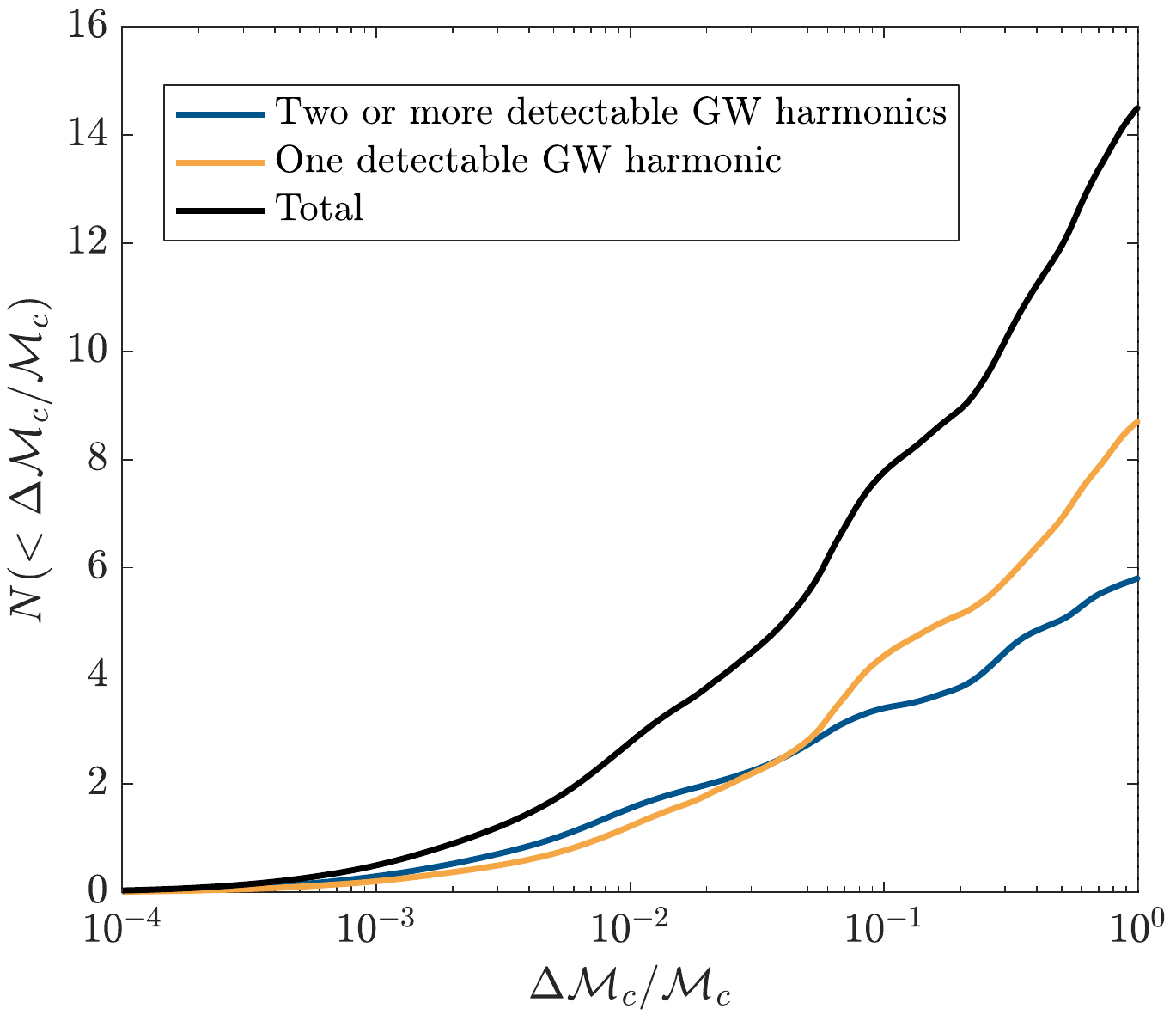}
	\caption{Cumulative distribution of chirp mass relative uncertainty for LISA \acp{DNS} (black), separated into those with one detectable \ac{GW} harmonic (orange) and with two or more detectable \ac{GW} harmonics (blue). We exclude sources with $\Delta \mathcal{M}_c / \mathcal{M}_c \geq 1$, for which LISA alone cannot measure the chirp mass.}
	\label{fig:mass_err}
\end{figure}

We plot the cumulative distribution of the chirp mass relative error of detectable LISA \acp{DNS} in Figure \ref{fig:mass_err}. For some sources, particularly low-frequency detections which do not appreciably evolve over the observation time (see Equations \ref{eq:fdot} and \ref{eq:Deltafdot}), the fractional chirp mass measurement uncertainty exceeds 1, meaning that LISA measurements alone cannot constrain the chirp mass. We exclude such sources from Figure \ref{fig:mass_err}. Among the $\approx 15$ \acp{DNS} with meaningful chirp mass constraints, those with two or more detectable harmonics only have marginally tighter mass constraints (median $\Delta \mathcal{M}_c / \mathcal{M}_c \approx 0.02$) than those with only one detectable harmonic (median $\Delta \mathcal{M}_c / \mathcal{M}_c \approx 0.05$). Although \acp{DNS} with only one detectable harmonic have poorer constrained absolute values of eccentricity (see Figure \ref{fig:ecc_err}), they tend to be less eccentric compared to sources with two detectable harmonics, and $\Delta F(e)/F(e) \propto e \Delta e$ for $e \rightarrow 0$, so the contribution of the eccentricity uncertainty to the chirp mass measurement error is small for low-eccentricity sources. A total of $\sim 8$ \acp{DNS} in our simulated population will have chirp masses constrained to better than 10 per cent in fractional uncertainty, which should be sufficient for the purpose of identifying the \ac{GW} source. The best-measured LISA \acp{DNS} will yield chirp masses with $\lesssim 1\%$ fractional uncertainty.

\section{Identifying a DNS with LISA} \label{sec:identifyingDNS}
Binary population synthesis studies estimate a population of $\sim10^8$ \acp{DWD} to exist in the \ac{MW} \citep[][and references therein]{Marsh2011}, most of which are expected to be detached \acp{DWD} \citep{Nelemans+2001b}. As discussed in Section \ref{subsec:SNR}, \acp{GW} emitted by unresolved Galactic \acp{DWD} form a confusion noise below 1--2 mHz, which has been included in the sensitivity curve used in this study \citep{Robson+2019}. However, $\sim 10^4$ binaries from this Galactic \acp{DWD} population are estimated to be detectable by LISA \citep{Farmer&Phinney2003,Nelemans+2001,Ruiter+2010,Korol+2017}, significantly outnumbering our estimated $\sim 30$ Galactic \acp{DNS}. Here, we discuss methods of positively identifying a \ac{DNS} with LISA observations.

The chirp mass is the primary means of differentiating \ac{DWD} and \ac{DNS} systems, with chirp masses above $\approx 1.2 \ \text{M}_\odot$ indicating that at least one component exceeds the Chandrasekhar limit for the maximum white dwarf mass. However, given the size of the \ac{DWD} population, a high-mass tail of $\mathcal{M}_c \lesssim 1.2 \ \text{M}_\odot$ (but sub-Chandrasekhar) \ac{DWD} binaries could still cause confusion with \acp{DNS}, as could neutron star-white dwarf binaries. 

The detection of a source with non-zero eccentricity favours a \ac{DNS} interpretation. The disc population of \acp{DWD} is thought to have formed via isolated binary evolution, where the progenitors are expected to have tidally-circularised from multiple mass transfer episodes \citep{Nelemans+2001b}. Observationally, there are no known eccentric Galactic \acp{DWD}, although there are observations of an eccentric Galactic pulsar-WD binary \citep{Antoniadis+2016} and a WD-main sequence \citep{Siess+2014} binary in the \ac{MW}. On the other hand, \acp{DNS} may have significant eccentricities from supernova and Blaauw kicks: in our model, half of LISA \acp{DNS} will have $e > 0.1$, and $\sim 10$ will have measurable second \ac{GW} harmonics, which allow eccentricity to be measured with $\Delta e \lesssim 0.02$ accuracy. Yet, dynamical formation channels in \ac{MW} globular clusters \citep{Willems+2007} or Lidov-Kozai oscillations in hierarchical triple systems \citep{Thompson2010} may produce eccentric \acp{DWD}. \citet{Kremer+2018} estimate that ejected binaries will only comprise a few \ac{MW} sources with $\rho \geq 2$, but given the very large \ac{DWD} population, even rare systems could be responsible for confusion with \acp{DNS}. 

The identification of an eccentric source as a \ac{DNS} is even more confident if a chirp mass measurement is possible. Above a chirp mass of $\approx 1.2 \ \text{M}_\odot$, the \ac{DWD} interpretation becomes highly unlikely. In fact, the chirp mass distribution of eccentric \acp{DWD} formed in globular clusters is expected to strong peak at 0.3--0.4$ \ \text{M}_\odot$ \citep{Willems+2007}.

Finally, sky localisation may also aid source identification. Since eccentric \acp{DWD} dynamically formed in \ac{MW} globular clusters are ejected into the Galactic halo, we expect eccentric disc binaries to be \acp{DNS}, though the latter may also be found far from the disc due to dynamical formation or kicks \citep[see, e.g., Figure C1 of][]{vignagomez2018}. Accurate sky localisation will also enhance the prospects for electromagnetic follow-up, which could definitively distinguish \ac{DNS} and \ac{DWD} systems \citep{Kyutoku+2019,Thrane+2019}.

\section{Conclusions \& Discussion} \label{sec:conclusions}
We estimated that around 35 inspiralling \acp{DNS} will be detectable over a four-year LISA mission with \ac{SNR} $\rho > 8$ using a mock population of isolated binaries synthesised with COMPAS. Of those, 94 per cent are expected to be Galactic \acp{DNS}, with the remainder in the LMC (5 per cent) and SMC (1 per cent). These \acp{DNS} are detected when the orbital frequency is typically 1 mHz, despite the presence of confusion-limited noise below \ac{GW} frequencies of 1--3 mHz from unresolved Galactic \ac{DWD} binaries.

Half of the detectable \acp{DNS} retain significant residual eccentricities, $e>0.11$, imparted mostly by the Blaauw kick at the second supernova in the COMPAS population synthesis models. Around a quarter of the LISA \acp{DNS} will have two or more individually detectable \ac{GW} harmonics and $\sim$ 40 per cent have only a single resolvable harmonic, while the remaining third will have \ac{GW} harmonics that combine to exceed the \ac{SNR} threshold, but are not individually resolvable. When two or more harmonics are observed, eccentricities may be accurately estimated to $\Delta e \lesssim 0.02$ by measuring \ac{SNR} ratios of different \ac{GW} harmonics. If only one \ac{GW} harmonic is observed for a \ac{DNS}, only an upper constraint on the eccentricity is placed at a typical level of $e \lesssim 0.1$.

A population of \acp{DNS} with well measured periods and eccentricities places valuable constraints on binary evolution physics. With a merger time of $\sim 2.4 \times 10^5$ years from a \ac{GW} frequency of $2f = 2$ mHz, the \acp{DNS} evolve slowly in frequency over the four year LISA mission, only changing their frequency by parts in $10^5$. This makes accurate chirp mass measurements challenging, which is compounded by the correlation between chirp mass and eccentricity in driving orbital frequency evolution. We find that $\approx 15$ \acp{DNS} will have useful chirp mass constraints from the LISA signal, with median fractional chirp mass uncertainties of $0.04$, dropping to below 1\% for the best-measured sources. These chirp mass and eccentricity measurements will make it possible to distinguish at least a fraction of the better-measured eccentric \acp{DNS} from the much larger Galactic \ac{DWD} population. They can also elucidate the origin of the \ac{DNS} systems: although the isolated binary channel is generally assumed to dominate \ac{DNS} formation, with globular clusters expected to contribute less than 10 per cent of all merging \acp{DNS} \citep{Phinney1991,Grindlay+2006,Ivanova+2008,Kremer+2018}, recent work has suggested that dynamical or three-body formation channels may be relevant \citep{Hamers&Thompson2019,Andrews&Mandel2019}. Moreover, LISA's measurement of the eccentricity distribution in the early \ac{DNS} evolutionary history could shed light on uncertainties in models of isolated binary evolution, such as the stability of case BB mass transfer.

LISA's heliocentric orbit produces an effective detector baseline of 2 AU for source triangulation, allowing for accurate sky localisation. We find that most \acp{DNS} will be localised with an angular resolution $\sigma_\theta \lesssim 2 \ \text{deg}$. This is sufficient to measure the height of Galactic \acp{DNS} relative to the Galactic plane to within $\sim 0.35$ kpc, which provides a constraint on the \ac{DNS} natal kick distribution. Around 6 \acp{DNS} will be localised sufficiently well to be covered by a single pointing of the \ac{SKA}, giving rise to an efficient, LISA-informed follow-up of possible radio pulsars.

The best-constrained LISA \acp{DNS}--the golden binaries--will be localised to a few arc-minutes with eccentricity inferred at an accuracy of a few parts in a thousand and the chirp mass to better than 1 per cent fractional uncertainty.

While this paper was under review, the manuscript of \cite{Andrews+2019} (hereafter A19) became available.  A19 study the population of LISA \acp{DNS} by sampling \ac{DNS} merger times and positions in the \ac{MW}.  They assume that all systems have periods and eccentricities set by the forward evolution of PSR B1913+16. They further assume a \ac{MW} merger rate of $210$ Myr${}^{-1}$ inferred from the \ac{DNS} \ac{GW} event GW170817 \citep{TheLIGOScientific:2017qsa}, which is $\approx 6$ times higher than our assumed rate of $33$ Myr${}^{-1}$ under our \texttt{Fiducial} model.  Therefore, A19 predict approximately 6 times more detections over a four-year LISA mission than we do. With the larger merger rate, A19 further predict $\sim 1$ detections in M31. A19 also find eccentricity uncertainties that are roughly consistent with ours, based on measuring the \ac{SNR} ratio of the $n=2,3$ harmonics. As shown in Figure \ref{fig:SNRratio}, for a typical LISA \ac{DNS}, the second and third harmonics have the highest \acp{SNR} only for $e\lesssim0.3$. For more eccentric \acp{DNS}, A19's approach overestimates the uncertainty.

\section*{Acknowledgements}
We thank Floor Broekgaarden, Philipp Podsiadlowski and Alberto Vecchio for discussions and suggestions. M. Y. M. L. acknowledges support by an Australian Government Research Training Program (RTP) Scholarship. A. V.-G. acknowledges funding support from Consejo Nacional de Ciencia y Tecnolog\'{\i}a (CONACYT). S. S. is supported by the Australian Research Council Centre of Excellence for Gravitational Wave Discovery (OzGrav), through project number CE170100004.






\bibliographystyle{mnras}
\bibliography{bibliography.bib}

\appendix
\section{Model variations in the \ac{DNS} frequency distribution}
\label{app:freq_dist_models}

\begin{figure}
	\centering
	\includegraphics[width=\columnwidth]{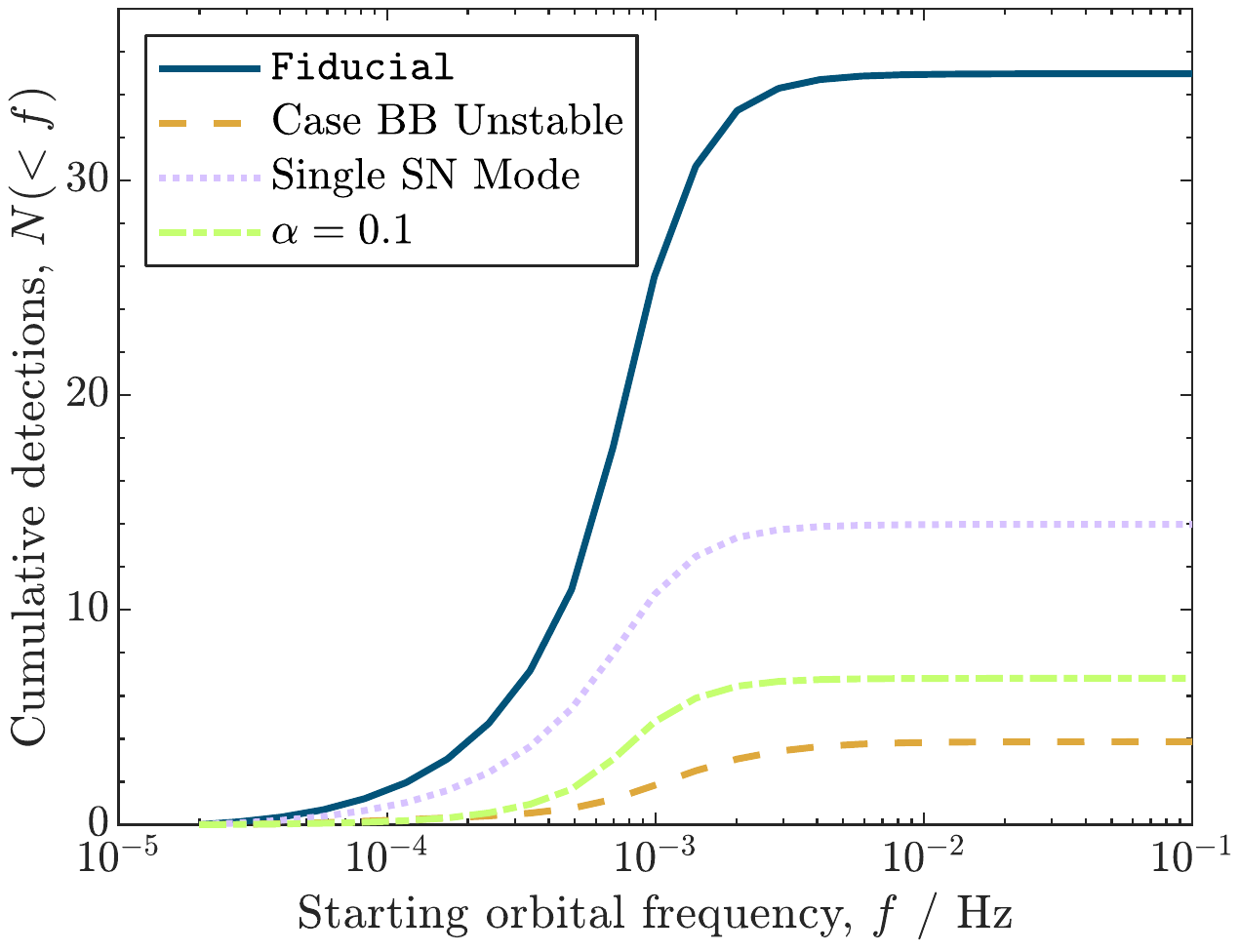}
	\caption{The cumulative number of LISA \ac{DNS} detections as a function of orbital frequency at the start of the observation for the COMPAS \texttt{Fiducial} model (solid blue curve) and for variations with: always dynamically unstable case BB mass transfer (orange dashed), a single \ac{SN} kick magnitude (purple dotted), and a common-envelope efficiency of $\alpha = 0.1$ (green dash-dot).}
	\label{fig:cumulativeDetectionsModelComparison}
\end{figure}

Figure \ref{fig:cumulativeDetectionsModelComparison} shows the distribution of the orbital frequencies of detectable \acp{DNS} at the start of LISA observation for the \texttt{Fiducial} model and the three model variations discussed in Section \ref{subsec:ecc}. The characteristic detection frequency is similar across the four models as it is mainly set by the LISA sensitivity.

The overall normalisation is, however, sensitive to changes in the binary evolution prescription. The \texttt{Fiducial} model yields the most \ac{DNS} detections among the considered variations. The single \ac{SN} mode causes fewer systems to be detected, as the higher natal kick scale parameter for ultra-stripped \acp{SN}, $\sigma_\text{high} = 265$ kms${}^{-1}$, compared to $\sigma_\text{low} = 30$ kms${}^{-1}$ in the \texttt{Fiducial} model, is more likely to disrupt the binary. The $\alpha = 0.1$ variation makes it approximately ten times more difficult to satisfy the global energy criterion for envelope ejection, $E_\text{bind} > \alpha E_\text{orb}$, compared to the \texttt{Fiducial} model where $\alpha = 1$, thereby decreasing the survivability of the common-envelope phase. On the other hand, the unstable case BB variation actually gives rise to a \ac{DNS} merger rate that is similar to the \texttt{Fiducial} model with stable case BB mass transfer, but with fewer detections because \acp{DNS} are produced at higher frequencies and quickly evolve through the LISA sensitivity window.


\bsp	
\label{lastpage}
\end{document}